\documentclass[aps,prl,superscriptaddress,showpacs,twocolumn,10pt]{revtex4-1}

\usepackage{graphicx}
\usepackage{amsmath,amssymb}
\usepackage{color}
\usepackage[colorlinks=true,linkcolor=blue,urlcolor=blue,citecolor=blue]{hyperref}
\usepackage{mathrsfs}
\usepackage[utf8]{inputenc}

\usepackage{bm}
\usepackage{physics}
\usepackage{mathtools}
\usepackage{mathptmx}
\usepackage{helvet}

\begin{document}

\title{Emission of Spin-correlated Matter-wave Jets from Spinor Bose-Einstein Condensates}
\author{Kyungtae Kim}
\affiliation{Department of Physics, Korea Advanced Institute of Science and Technology, Daejeon 34141, Korea }
\author{Junhyeok Hur}
\affiliation{Department of Physics, Korea Advanced Institute of Science and Technology, Daejeon 34141, Korea }
\author{SeungJung Huh}
\affiliation{Department of Physics, Korea Advanced Institute of Science and Technology, Daejeon 34141, Korea }
\author{Soonwon Choi}
\affiliation{Department of Physics, University of California Berkeley, Berkeley, CA 94720, USA}
\affiliation{Center for Theoretical Physics, Massachusetts Institute of Technology, Cambridge, MA 02139, USA}
\author{Jae-yoon Choi}
\email[]{jaeyoon.choi@kaist.ac.kr}
\affiliation{Department of Physics, Korea Advanced Institute of Science and Technology, Daejeon 34141, Korea }

\date{\today}

\begin{abstract}

We report the observation of matter-wave jet emission in a strongly ferromagnetic spinor Bose-Einstein condensate of $^7$Li atoms. Directional atomic beams with $\ket{F=1,m_F=1}$ and $\ket{F=1,m_F=-1}$ spin states are generated from $\ket{F=1,m_F=0}$ state condensates, or vice versa. This results from collective spin-mixing scattering events, where spontaneously produced pairs of atoms with opposite momentum facilitates additional spin-mixing collisions as they pass through the condensates. The matter-wave jets of different spin states ($\ket{F=1,m_F=\pm1}$) can be a macroscopic Einstein-Podolsky-Rosen state with spacelike separation. Its spin-momentum correlations are studied by using the angular correlation function for each spin state. Rotating the spin axis, the inter-spin and intra-spin momentum correlation peaks display a high contrast oscillation, indicating collective coherence of the atomic ensembles. We provide numerical calculations that describe the experimental results at a quantitative level and can identify its entanglement after 100~ms of a long time-of-flight.

\end{abstract}
\maketitle

The collective scattering processes in many-body systems can lead to remarkable, counter-intuitive phenomena, due to quantum interference effects. Superradiance in an atomic ensemble is a prominent example, where the spontaneous emission process occurs cooperatively, emitting a directional light with an enhanced decay rate~\cite{Dicke1954}. Superradiant scattering is observed in degenerate Bose~\cite{Inouye1999} and Fermi gases~\cite{Pengjun2011}, and is often described as a self-amplified atom-light scattering process~\cite{Inouye1999,Moore1999,Vardi2002}. Atoms scattered by incoming light can interfere with condensates at rest, forming a matter-wave grating that diffracts successive laser light, enhancing the amplitude of the density modulation. These collective behaviors are not solely restricted to the optical domain, but can also extend to matter waves, where directional atomic beams are generated without external light fields~\cite{Clark2017}. Under a periodic driving of scattering length, density modulations are spontaneously developed, stimulating further pair-wise collision processes with a certain direction given by the density modulation~\cite{Clark2017, Fu2018}. This ends up as a directional atomic beam, resembling fireworks, contrasting a diffusive spherical shell structure out of uncorrelated $s$-wave collisions~\cite{Chikkatur2000}.

Recent studies of matter-wave emissions have offered new opportunities to study complex correlations in the high-harmonic generation process~\cite{Feng2019}, quantum phenomena in a non-inertia frame~\cite{Hu2019}, and dissipative many-body quantum dynamics~\cite{Krinner2018}. Such efforts can provide new directions for producing non-classical quantum states of atomic spins. In spinor Bose-Einstein condensates (BECs), for example, correlated spin states like the squeezed vacuum state~\cite{Gross2011,Lucke2011,Hamley2012} have been generated via spin-mixing collisions, to explore fundamental questions of quantum physics~\cite{Lange2018, Kunkel2018} and its application to quantum metrology~\cite{Muessel2014, Kruse2016,Pezze2018}. However, most of such experiments have focused on the low kinetic energy regime, where the created spin pairs are localized in the trapping potential with their source, challenging the local addressing and manipulation of the quantum state. One wave to overcome this hurdle would be to realize directional superradiant collisions in a spin manifold, which could generate a macroscopic Einstein-Podolsky-Rosen (EPR) state of atoms~\cite{Einstein1935,Pu2000,Duan2000}. 

In this Letter, we report the emission of spin-correlated matter-wave jets from spinor BECs of $^7$Li atoms. The directional atomic beams of $\ket{F=1,m_F=\pm1}(\ket{\uparrow},\ket{\downarrow})$ spin states are generated from $\ket{F=1,m_F=0}$ state condensates, or vice versa. The kinetic energy of the atomic beam is high enough to escape the trapping potential, where we take advantage of strongly ferromagnetic spin interaction to facilitate the matter-wave amplification of fast-moving particles. The matter-waves with opposite spin states can be a macroscopic EPR state, and its spin-momentum correlation is revealed by angular correlation functions between emitted spin states. To investigate non-classical correlations, we coherently rotate the spin states and study the responses of the correlation functions at various rotation angles. The momentum correlation peaks among these spin states exhibit a high contrast oscillation as a function of the rotation angle, suggesting that the ensembles of atoms with opposite spins still maintain collective coherence to exhibit interference patterns. Numerical analysis is provided indicating the non-classical correlations can be observed after a sufficiently long expansion time ($\sim$100~ms).

\begin{figure*}
\centering
\includegraphics[width=1\linewidth]{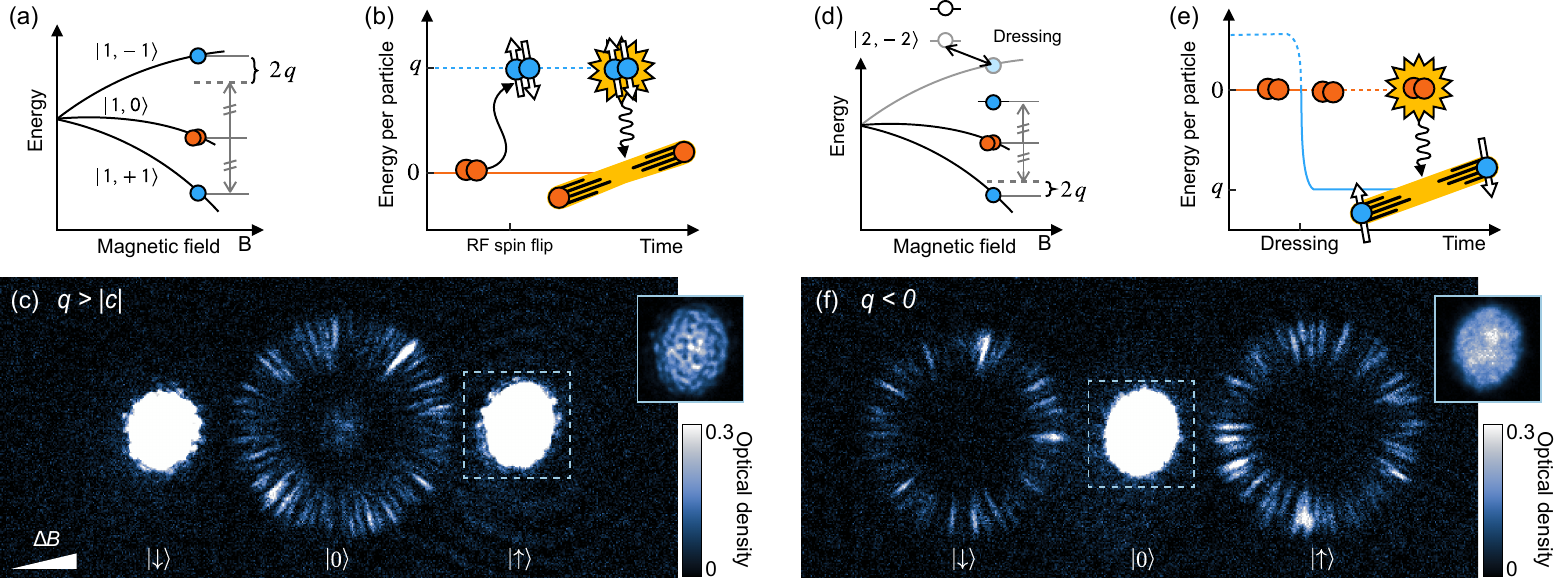}
\caption{Emission of matter-wave jets from spinor BECs.
(a) Zeeman shift of the $F = 1$ ground state manifold under a magnetic field $B$. The hyperfine interaction introduces a quadratic Zeeman shift $q>0$ for $^7$Li atoms. 
(b) An RF pulse flips atoms in the $\ket{0}$ state to the $(\ket{\uparrow}-\ket{\downarrow})/\sqrt{2}$ state, which can be a highly excited spin state under a magnetic field ($q>|c|$). Quantum fluctuations produce two atom pairs of $\ket{0}$ state after spin-mixing collisions, and the excess internal energy is released in the kinetic energy of the atoms, flying in opposite directions because of momentum conservation.
(c) Spin-resolved image of the matter-wave jets in $\ket{0}$ state with $q$ = 1.7 kHz after a hold time of $t_h=7.5$~ms. The image displays atoms in the horizontal plane, where we apply a field gradient during 16~ms of the time-of-flight (TOF). From left to right, the corresponding spins are $\{\ket{\downarrow}, \ket{0}, \ket{\uparrow}\}$. The inset represents the condensates remaining in the trap (dashed box). 
(d) Dressing the $\ket{F=1, m_z=-1}$ state with $\ket{F=2, m_z=-2}$ state by applying a microwave, we tune the $q$ to have a negative value~\cite{Gerbier2006}. 
(e) After switching on the dressing field, the polar phase is no longer the ground state of the system, creating spin pairs ($\ket{\uparrow}$ and $\ket{\downarrow}$). Through the de-excitation process, the atom obtains kinetic energy from $|q|$. (f) After sufficient stimulated collisions, matter-wave jets of opposite spins are also observed. 
}
\label{fig1}
\end{figure*}

Our experimental sequences start by preparing an $F=1$ spinor condensate of $^7$Li atoms in a quasi-two-dimensional (2D) optical trap~\cite{Huh2020}. The condensates are initially in a polar phase due to a few Gauss of the magnetic field along the $z$-axis. To observe the matter-wave jets from spinor BECs, we apply an RF pulse to the trapped condensates, making equal population of $\ket{\uparrow}$ and $\ket{\downarrow}$ states. When the quadratic Zeeman energy, $q=h\times1-5$~kHz [Fig.~1(a)], dominates the spin-dependent interaction energy ($|c|=h\times160$~Hz), the initial state is unstable, producing pairs of atom in the $\ket{0}$ state owing to the presence of quantum fluctuations~\cite{Kawaguchi2012}. The spontaneously created atom pairs obtain kinetic energy from the quadratic Zeeman energy and propagate in opposite direction due to momentum conservation [Fig.~1(b)]. The moving atoms can interfere with stationary condensates in the $\ket{\uparrow}$ and $\ket{\downarrow}$ states via spin-mixing Hamiltonian, displaying spatial modulations in the scattering amplitude that stimulates a further pair generation process~\cite{Pu2000, Vardi2002}. This, in turn, leads to self-amplification of the modulation amplitude and the emission of a directional atomic beam in the horizontal plane constrained by strong 2D potential~\cite{SM2021}.

The spin-mixing collisional strength is characterized by the spin-dependent interaction coefficient ($c_2$). We note that the $^7$Li atoms are favorable for observing the matter-wave jets because of their strong spin interactions~\cite{Huh2020}. The spin-dependent interaction coefficient of the atoms is as large as 46$\%$ of the spin-independent interaction coefficient. Therefore, the grating formed by spin-mixing interaction is weakly dephased more by the source condensates than the other alkali atoms~\cite{Vogels2002}, and the traveling spin pairs can be amplified self-consistently.

The emission of matter-wave jets in the $\ket{0}$ state is shown in Fig.~1(c). Stern-Gerlach spin separated absorption images are used to resolve the source condensates in the trap and the created pairs, so that we are able to study its dynamics under various hold times $t_h$ after the RF pulse. In the first few milliseconds of hold time, the condensates are stable with no populations in the $\ket{0}$ state. Then, radially propagating atomic beams with narrow angular width, matter-wave jets, suddenly appear at $t_h\sim5$~ms. After 10~ms of hold time, the matter-wave jets start to escape the BECs. The angular pattern of the jets is random in each experimental run, and the jets seem to have their own partner in the opposite direction. Moreover, we find that the kinetic energy per atom ($E_k$) of the matter-wave jets is almost equal to the quadratic Zeeman energy~\cite{SM2021}, indicating that the atom pairs are created from the source condensates after the spin-changing scattering process. Since the quadratic Zeeman energy far exceeds the condensate chemical potential, $\mu=h\times300$~Hz, the matter-wave jets can escape the BECs and trapping potential.

As a complementary experiment, we also investigate the emission of matter-wave jets with $\ket{\uparrow}$ and $\ket{\downarrow}$ states from polar condensates. The polar phase becomes dynamically unstable when the quadratic Zeeman energy is negative~\cite{Kawaguchi2012}, generating spin pairs of the $\ket{\uparrow}$ and $\ket{\downarrow}$ states [Fig.~1(e)]. Similar to the previous experiment, we observe two-dimensional matter-wave jets of the created spin states that have kinetic energy from the  quadratic Zeeman energy, $E_k\simeq|q|$. One remarkable difference is that the matter-wave jets can naturally have a spin-momentum correlation (reminding the EPR state), which will be discussed after studying its generation mechanism.

\begin{figure}
\centering
\includegraphics[width=0.8\linewidth]{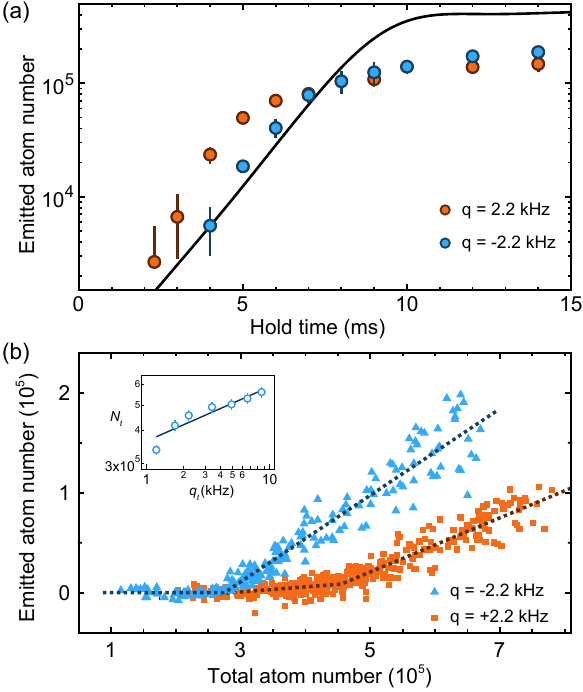}
\caption{(a) Dynamical generation of matter-wave jets. Emitted atom numbers over hold time for $q=\pm2.2$~kHz. The error bars show one standard deviation (SD). The solid line is obtained by numerical calculation based on the Bogoliubov theory of spinor condensate with the effective growth rate $\gamma_e=160$~Hz~\cite{SM2021}. (b) Threshold for jet formation. Emitted atom number after $t_h=15$~ms under various initial atom numbers. To extract the threshold atom number, $N_t$, the data is fitted to bilinear curves (dashed lines). Inset shows $q_t$ dependence of $N_t$ with power-law fit, $N_t\sim q_t^{0.2}$ (solid line). The error bars mean 95$\%$ confidence interval of the bilinear fits.
}
\label{fig2}
\end{figure}

The creation process can be well understood by investigating the early time dynamics of matter-wave jets [Fig.~2(a)]. For both experiments listed above, initial dynamics are well described by a simple exponential function, $N_{jet}(t)=N_{jet}(0)e^{\gamma_e t}$, where the $\gamma_e\sim160$~Hz. In the long-time limit, the jet populations saturate because of depletion of the source. The exponential growth dynamics is a characteristic feature of the dynamical instability, in which spontaneously created atoms pairs are parametrically amplified~\cite{Klempt2010}. The microscopic origin of such instability can be found in the existence of imaginary eigenfrequencies in the Bogoliubov quasi-particle (atom pair) spectrums~\cite{Kawaguchi2012}. The instability rate is maximized for particles with kinetic energy $E_k=|q|-c$, which is consistent with the observation within measurement uncertainty. The population growth rate of the quasi-particles is proportional to the spin-dependent interaction energy, $\gamma_b=2|c|/\hbar=320$~Hz, which is two times higher than the observation. In order to ascribe the discrepancy, we consider a loss rate $\kappa$ in a finite-sized system. When atoms with velocity $v$ leave the condensates of radius $R_{\text{TF}}$ before sufficient spin-mixing collisions, it leads to an atom loss with a rate $\kappa\sim v/R_{\text{TF}}$~\cite{Vardi2002, Clark2017}. In the experiments, we have $R_{\text{TF}}=100~\mu{}$m and $|q|=h\times2.2$~kHz, and the escape rate $(v/R_{\text{TF}})$ is evaluated to 100~Hz, which may account for the observed difference between $\gamma_e$ and $\gamma_b$.

From this competing relation, we can expect that the burst mode only occurs when $\gamma_b>\kappa$. In other words, for runaway stimulated collisions certain thresholds of atom number ($N>N_t$) and quadratic Zeeman energy ($q<q_t$) are required. For both initial conditions ($q/h=2.2$~kHz and $q/h=-2.2$~kHz), we observe a clear threshold behavior [Fig.~2(b)], and the threshold atom number increases with the external magnetic field [Fig.~2(b) inset]. An interesting observation is that even with the same condensate density and magnitude of the quadratic Zeeman energy $|q|$, the threshold atom number $N_t$ for negative $q$ is smaller than that of positive $q$. We attribute such difference to the immiscible dynamics between the $\ket{\uparrow}$ and $\ket{\downarrow}$ states~\cite{De2014}, which form magnetic spin domains after a hold time [Fig.~1(c) inset]. We suspect phase diffusion dynamics may occur during the domain wall formation, which could suppress the collective pair production process for positive $q$.


\begin{figure}
\centering
\includegraphics[width=0.9\linewidth]{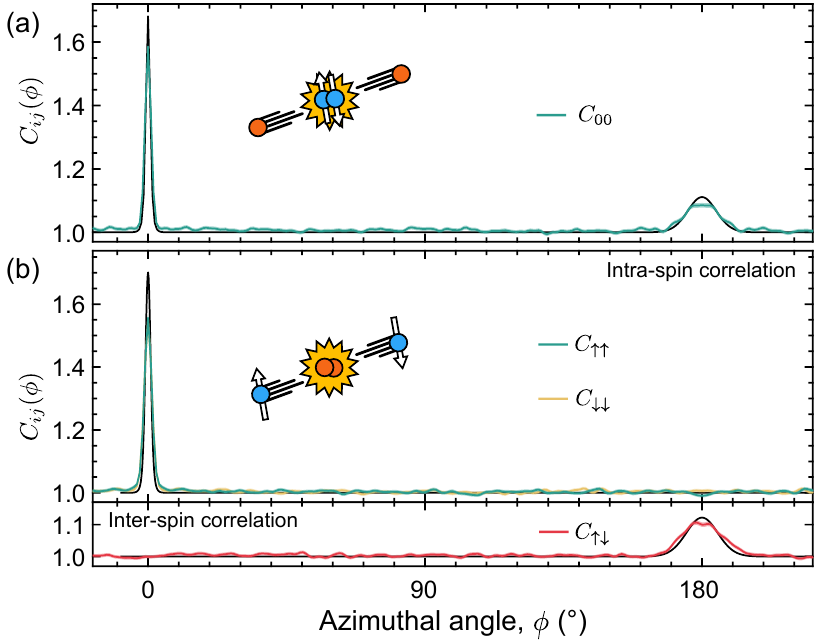}
\caption{Angular correlation functions of matter-wave jets at after 28~ms of TOF under (a) $q$ = 2.7 kHz and (b) $q$ = -2 kHz. (b) The upper panel shows the correlation between the same spins (intra-spin) and the lower panel shows the correlation between the opposite spins (inter-spin). Shaded areas mark one standard error based on 100 to 120 realizations. Black solid lines are time-dependent the Bogoliubov calculations for the experimental parameters.
}
\label{fig3}
\end{figure}

To uncover correlations in the matter-wave jets, we study the angular correlation functions.
\begin{equation}
C_{ij}(\phi) = \frac{\left\langle n_{i}(\phi') n_{j}(\phi'+\phi) \right\rangle}{\left\langle n_{i}(\phi')\right\rangle \left\langle n_{j}(\phi'+\phi) \right\rangle},
\end{equation} where $n_i(\phi)$ is the angular density of the emitted atoms in the $\ket{i}$ state and the brackets mean the angular and ensemble average, $\ev{n_i}  = \ev{\frac{1}{2\pi}\int{ n_i(\phi') \ d\phi'}}_{\text{ens}}$. When the jets are in the $|0\rangle$ state for $q>|c|$, we study the correlation function $C_{00}(\phi)$, and when jets are in $\ket{\uparrow}$ and $\ket{\downarrow}$ states, we study spin versus angular-moment correlations by investigating the inter-spin $C_{\uparrow\downarrow}(\phi)$ and intra-spin states $C_{\uparrow\uparrow}(\phi)$ $\&$ $C_{\downarrow\downarrow}(\phi)$ correlation functions. For both experiments, we observe one sharp peak near $\phi=0^{\circ}$ and the other broad peak near $\phi=180^{\circ}$. Both peaks are well displayed in $C_{00}$ [Fig.~3(a)], and for matter-wave jets of opposite spin states, each peak can be found in the intra- and inter-spin state correlation functions [Fig.~3(b)], respectively.

The peak near $\phi=0^{\circ}$ is a result of the Hanbury Brown-Twiss (HBT) effect between different angular modes, which would be close to 2 in the ideal limit~\cite{Hanbury1956,Wu2019}. The variance of $\langle N_k\rangle$ (atom number with momentum $k$) becomes $\langle N_k\rangle[\langle N_k\rangle+1]$, since $\sigma^2=[C_{ii}(0)-1]\langle N_k\rangle^2+\langle N_k\rangle$ and $C_{ii}(0)=2$. This result indicates thermal-like fluctuations of the emitted jet populations. As indicated in the study~\cite{Mias2008}, the reduced density matrix of the quantum state displays a Bose-Einstein distribution because of the entropy associated with the correlated pairs. Indeed, the angular mode population for both spin states is well described by the thermal distribution~\cite{SM2021}. A similar observation is also reported from the Ref.~\cite{Hu2019}, where the authors connect the emerging thermal distribution to the Unruh radiation, and the pair production Hamiltonian can be considered as a boosting transformation in an accelerating frame. 

The other peak near $\phi=180^{\circ}$ implies momentum conservation of the emitted atom pair, and its width is broadened because of the near-field effect~\cite{Fu2018,Wu2019}. That is, the condensates cannot be regarded as a point source with the current expansion time so that different modes are overlapped in the detection area and interfere. According to the simple near-field model~\cite{Fu2018}, about three modes ($N_m\simeq3$) can be overlapped, reducing the correlation peaks to $C_{ij}(\phi=180^{\circ})= 1+1/N_m^2\simeq 1.1$. This is further supported by our time-dependent Bogoliubov calculation, which captures the angular correlation functions for all spin states at a quantitative level [Fig.~3(a) and (b)]. As we increase the expansion time in the calculation, these modes can be resolved, and the peak becomes higher and narrower.

\begin{figure}
\centering
\includegraphics[width=0.9\linewidth]{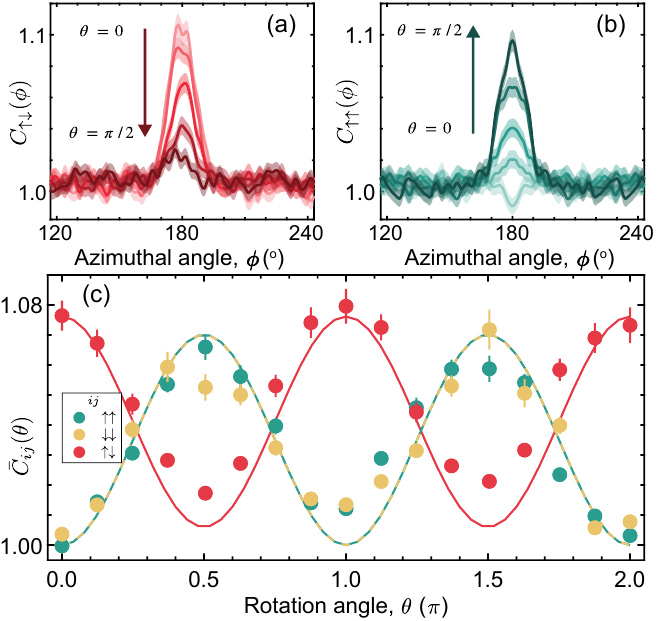}
\caption{As the spin rotation angle $\theta$ is increased from $0$ to $\pi/2$, (a) the inter-spin correlation decreases (from light red to dark red) whereas, (b) the intra-spin correlation increases (from light green to dark green). The shades represent one standard deviation of the mean over 110 independent experiments. (c) Evolution of the averaged momentum correlation peak $\bar{C}_{ij}$ (mean value of $C_{ij}(\phi)$ in the interval $\phi\in[170^{\circ}, 190^{\circ}]$) under spin-axis rotation. Solid and dashed lines are numerical calculations accounting the field gradient effect \cite{SM2021}. The error bars represent one standard deviation of the mean.
}
\label{fig4}
\end{figure}

Matter-wave jets of opposite spins are of particular interest as the momentum correlation peak only appears for inter-spin state, $C_{\uparrow\downarrow}(\phi=180^{\circ})$, indicating strongly correlation between spins and momentum like EPR state. In a homogeneous system (or infinite expansion time limit), the quantum state after time evolution can be written as,
\begin{equation} \label{eq:jet} 
\bigotimes _{ \epsilon_{\boldsymbol{k}} \approx |q| }\sum_{j=0}^{\infty}\lambda_{j}(t)\left[\hat{a}_{ \boldsymbol{k},\uparrow}^{\dagger}\hat{a}_{ \boldsymbol{-k},\downarrow}^{\dagger} + \hat{a}_{ \boldsymbol{k},\downarrow}^{\dagger}\hat{a}_{ \boldsymbol{-k},\uparrow}^{\dagger}\right]^j \ket{\text{vac}}.
 \end{equation} 
The tensor product excludes opposite momentum ($\boldsymbol{k}=-\boldsymbol{k}$), and $\lambda_{j}(t) = \frac{{(-i\tanh{(c t)})^j}}{j!\cosh^2{(ct)}}$~\cite{SM2021}. Here, the triplet Bell pair state $\ket{\Psi_{T}}=(\hat{a}_{k,\uparrow}\hat{a}_{-k,\downarrow}+\hat{a}_{k,\downarrow}\hat{a}_{-k,\uparrow})^{\dagger}\ket{\text{vac}}$ constitutes the macroscopic entangled state, whose non-classical correlation can be shown by interfering the two spin states. For example, under a spin rotation ($\ket{\uparrow},\ket{\downarrow}\rightarrow\frac{\ket{\uparrow}\pm\ket{\downarrow}}{\sqrt{2}}$), the triplet state becomes another entangled state, $\ket{\Psi_{+}}=(\hat{a}_{k,\uparrow}\hat{a}_{-k,\uparrow}+\hat{a}_{k,\downarrow}\hat{a}_{-k,\downarrow})^{\dagger}\ket{\text{vac}}$. Analyzing the angular correlation function $C_{ij}(\phi)$ for the entangled state $\ket{\Psi_{+}}$, the correlation peak near $\phi=180^{\circ}$ would appear only for the same spin state ($C_{\uparrow\uparrow}~\&~C_{\downarrow\downarrow}$). Increasing the spin rotation angle $\theta$  from 0 to $2\pi$, these two quantum states are transformed into each other, displaying oscillations in the intra- and inter-spin state correlation peaks at $\phi=180^{\circ}$.

Taking the $|F=2,m_F=0\rangle$ upper hyperfine state as an intermediate state, the spin axis can be rotated by applying a microwave pulse~\cite{Lucke2011}. When the matter-wave jets start to escape the trapped BECs ($t_h=7.5$~ms), we apply the rotating pulse and measure the correlation functions. The momentum correlation peak in $C_{\uparrow\downarrow}(\phi=180^{\circ})$ gradually disappears [Fig.~4(a)], while it simultaneously emerges in the intra-spin correlation functions $C_{\uparrow\uparrow}$ and $C_{\downarrow\downarrow}$ [Fig.~4(b)]. Displaying the momentum-correlation peaks in Fig.~4(c) as a function of the rotation angle, we observe a clear oscillation with almost full contrast after $2\pi$ rotation.

The observed oscillations are clear signature that our spin states are in a coherent superposition state with spin-momentum correlations. However, this does not constitute the witness of macroscopic entanglement in a single-mode, as each jet stream contains many particles in distinct modes, and the global wavefunction may not be pure. Searching for non-classical correlations, we take a bipartite separability criterion for the collective spins~\cite{Iskhakov2012}.
\begin{equation}
        \textsf{W} =\sum_{i=x, y, z} \frac{\text{Var}\qty(\hat{\text{J}}_i)}{2\ev{\hat{\text{J}}_0}},
\end{equation}
where $\textsf{W}<1$ implies that the density matrix of the system cannot be described by classical correlations, certifying the presence of quantum entanglement. Here, the collective spin vector operators and the normalization factor are defined as, 
\begin{eqnarray} \label{eq:witness}
	\nonumber
        \ev{\hat{\text{J}}_0} &=& \ev{\sum_{m=\uparrow,\downarrow}[N_m(\phi) + N_m(\phi+\pi) ]},\\ 
        \nonumber
        \hat{\text{J}}_i &=& \hat{P}_i(\phi) \pm  \hat{P}_i(\phi+\pi),\quad \\ 
        \nonumber
        \hat{P}_i(\phi) &=& \int^{\Delta\phi/2}_{-\Delta\phi/2}{ [ \hat{n}_{\uparrow i}(\phi+\phi') - \hat{n}_{\downarrow i}(\phi+\phi') ]d\phi' }.
\end{eqnarray} The index $i=(x,y,z)$ refers to the spin axis, and the transverse (longitudinal) spin length takes a minus (plus) sign.

Figure~5 displays the experimentally measured and theoretically calculated entanglement witness as a function of the angular bin size $\Delta\phi$. We assume the transverse spin length to be the averaged value of the $\hat{\text{J}}_x$ and $\hat{\text{J}}_y$ because the phase of the spin state is independent of the microwave phase. Both results show that the entanglement witness is far above the entanglement criterion $\textsf{W}<1$. The experiment's deviation from the calculation is attributed to the residual field gradient effect, atom loss during the microwave dressing, and atom number counting uncertainties~\cite{SM2021}. Even without these experimental artifacts, however, the calculation shows that the interference between different angular momentum modes dims the information of the corresponding pair, so that the non-separability criterion cannot be satisfied with 28~ms of expansion time. This trap geometry requires a long expansion time ($\sim$100~ms) to resolve individual angular momentum modes, where we observe the two correlation peaks become almost symmetric. In this respect, reducing the possible number of emitting modes can be a promising way to generate certifiable macroscopic entanglement, for example, by using quasi-one-dimensional trap geometry that exhibits only two outgoing modes~\cite{Meznarsic2020}. 


\begin{figure}
\centering
\includegraphics[width=0.9\linewidth]{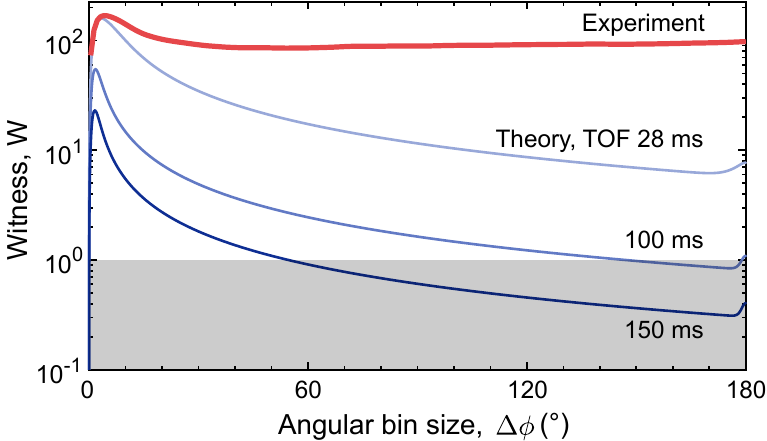}
\caption{Entanglement witness at various angular bin sizes. The entanglement can be certified when $\textsf{W}<1$ (gray area). The experiment (red line) and Bogoliubov calculation (light blue) show that many angular momentum modes are integrated on a single bin with 28~ms of expansion time, losing the information of the counterpart. Only with sufficiently long TOF (dark blue line), we are able to claim the non-classical correlations in the matter-wave jets
}
\label{fig5}
\end{figure}

In conclusion, we have observed directional matter-wave jets of various spin states via the {\textquotedblleft}superradiant{\textquotedblright} spin-mixing scattering process in spinor condensates. The matter-wave jets exhibit strong correlations between spins and momentum and hold the promise of being a macroscopic EPR state. This work opens up new perspectives for the study of quantum atom optics and quantum simulations using spinor BECs. It can be applied to precision measurement in an atom interferometer~\cite{Pezze2018}, to test Bell's inequality with massive particles~\cite{Bell1964}, and to study non-equilibrium quantum dynamics with fast-moving impurities in a quantum liquid~\cite{Mathy2012}. 

The authors thank Andrey Moskalenko and Young-Sik Ra for discussion. This work was supported by the National Research Foundation of Korea Grant No.2019M3E4A1080401 and Samsung Science and Technology Foundation BA1702-06. S.C. acknowledges support from the Miller Institute for Basic Research in Science. 

\nocite{key}
%


\begin{thebibliography}{45}%
\makeatletter
\providecommand \@ifxundefined [1]{%
 \@ifx{#1\undefined}
}%
\providecommand \@ifnum [1]{%
 \ifnum #1\expandafter \@firstoftwo
 \else \expandafter \@secondoftwo
 \fi
}%
\providecommand \@ifx [1]{%
 \ifx #1\expandafter \@firstoftwo
 \else \expandafter \@secondoftwo
 \fi
}%
\providecommand \natexlab [1]{#1}%
\providecommand \enquote  [1]{``#1''}%
\providecommand \bibnamefont  [1]{#1}%
\providecommand \bibfnamefont [1]{#1}%
\providecommand \citenamefont [1]{#1}%
\providecommand \href@noop [0]{\@secondoftwo}%
\providecommand \href [0]{\begingroup \@sanitize@url \@href}%
\providecommand \@href[1]{\@@startlink{#1}\@@href}%
\providecommand \@@href[1]{\endgroup#1\@@endlink}%
\providecommand \@sanitize@url [0]{\catcode `\\12\catcode `\$12\catcode
  `\&12\catcode `\#12\catcode `\^12\catcode `\_12\catcode `\%12\relax}%
\providecommand \@@startlink[1]{}%
\providecommand \@@endlink[0]{}%
\providecommand \url  [0]{\begingroup\@sanitize@url \@url }%
\providecommand \@url [1]{\endgroup\@href {#1}{\urlprefix }}%
\providecommand \urlprefix  [0]{URL }%
\providecommand \Eprint [0]{\href }%
\providecommand \doibase [0]{http://dx.doi.org/}%
\providecommand \selectlanguage [0]{\@gobble}%
\providecommand \bibinfo  [0]{\@secondoftwo}%
\providecommand \bibfield  [0]{\@secondoftwo}%
\providecommand \translation [1]{[#1]}%
\providecommand \BibitemOpen [0]{}%
\providecommand \bibitemStop [0]{}%
\providecommand \bibitemNoStop [0]{.\EOS\space}%
\providecommand \EOS [0]{\spacefactor3000\relax}%
\providecommand \BibitemShut  [1]{\csname bibitem#1\endcsname}%
\let\auto@bib@innerbib\@empty
\bibitem [{\citenamefont {Dicke}(1954)}]{Dicke1954}%
  \BibitemOpen
  \bibfield  {author} {\bibinfo {author} {\bibfnamefont {R.~H.}\ \bibnamefont
  {Dicke}},\ }\href {\doibase 10.1103/PhysRev.93.99} {\bibfield  {journal}
  {\bibinfo  {journal} {Phys. Rev.}\ }\textbf {\bibinfo {volume} {93}},\
  \bibinfo {pages} {99} (\bibinfo {year} {1954})}\BibitemShut {NoStop}%
\bibitem [{\citenamefont {Inouye}\ \emph {et~al.}(1999)\citenamefont {Inouye},
  \citenamefont {Chikkatur}, \citenamefont {Stamper-Kurn}, \citenamefont
  {Stenger}, \citenamefont {Pritchard},\ and\ \citenamefont
  {Ketterle}}]{Inouye1999}%
  \BibitemOpen
  \bibfield  {author} {\bibinfo {author} {\bibfnamefont {S.}~\bibnamefont
  {Inouye}}, \bibinfo {author} {\bibfnamefont {A.~P.}\ \bibnamefont
  {Chikkatur}}, \bibinfo {author} {\bibfnamefont {D.~M.}\ \bibnamefont
  {Stamper-Kurn}}, \bibinfo {author} {\bibfnamefont {J.}~\bibnamefont
  {Stenger}}, \bibinfo {author} {\bibfnamefont {D.~E.}\ \bibnamefont
  {Pritchard}}, \ and\ \bibinfo {author} {\bibfnamefont {W.}~\bibnamefont
  {Ketterle}},\ }\href {\doibase 10.1126/science.285.5427.571} {\bibfield
  {journal} {\bibinfo  {journal} {Science}\ }\textbf {\bibinfo {volume}
  {285}},\ \bibinfo {pages} {571} (\bibinfo {year} {1999})}\BibitemShut
  {NoStop}%
\bibitem [{\citenamefont {Wang}\ \emph {et~al.}(2011)\citenamefont {Wang},
  \citenamefont {Deng}, \citenamefont {Hagley}, \citenamefont {Fu},
  \citenamefont {Chai},\ and\ \citenamefont {Zhang}}]{Pengjun2011}%
  \BibitemOpen
  \bibfield  {author} {\bibinfo {author} {\bibfnamefont {P.}~\bibnamefont
  {Wang}}, \bibinfo {author} {\bibfnamefont {L.}~\bibnamefont {Deng}}, \bibinfo
  {author} {\bibfnamefont {E.~W.}\ \bibnamefont {Hagley}}, \bibinfo {author}
  {\bibfnamefont {Z.}~\bibnamefont {Fu}}, \bibinfo {author} {\bibfnamefont
  {S.}~\bibnamefont {Chai}}, \ and\ \bibinfo {author} {\bibfnamefont
  {J.}~\bibnamefont {Zhang}},\ }\href {\doibase 10.1103/PhysRevLett.106.210401}
  {\bibfield  {journal} {\bibinfo  {journal} {Phys. Rev. Lett.}\ }\textbf
  {\bibinfo {volume} {106}},\ \bibinfo {pages} {210401} (\bibinfo {year}
  {2011})}\BibitemShut {NoStop}%
\bibitem [{\citenamefont {Moore}\ and\ \citenamefont
  {Meystre}(1999)}]{Moore1999}%
  \BibitemOpen
  \bibfield  {author} {\bibinfo {author} {\bibfnamefont {M.~G.}\ \bibnamefont
  {Moore}}\ and\ \bibinfo {author} {\bibfnamefont {P.}~\bibnamefont
  {Meystre}},\ }\href {\doibase 10.1103/PhysRevLett.83.5202} {\bibfield
  {journal} {\bibinfo  {journal} {Phys. Rev. Lett.}\ }\textbf {\bibinfo
  {volume} {83}},\ \bibinfo {pages} {5202} (\bibinfo {year}
  {1999})}\BibitemShut {NoStop}%
\bibitem [{\citenamefont {Vardi}\ and\ \citenamefont
  {Moore}(2002)}]{Vardi2002}%
  \BibitemOpen
  \bibfield  {author} {\bibinfo {author} {\bibfnamefont {A.}~\bibnamefont
  {Vardi}}\ and\ \bibinfo {author} {\bibfnamefont {M.~G.}\ \bibnamefont
  {Moore}},\ }\href {\doibase 10.1103/PhysRevLett.89.090403} {\bibfield
  {journal} {\bibinfo  {journal} {Phys. Rev. Lett.}\ }\textbf {\bibinfo
  {volume} {89}},\ \bibinfo {pages} {090403} (\bibinfo {year}
  {2002})}\BibitemShut {NoStop}%
\bibitem [{\citenamefont {Clark}\ \emph {et~al.}(2017)\citenamefont {Clark},
  \citenamefont {Gaj}, \citenamefont {Feng},\ and\ \citenamefont
  {Chin}}]{Clark2017}%
  \BibitemOpen
  \bibfield  {author} {\bibinfo {author} {\bibfnamefont {L.~W.}\ \bibnamefont
  {Clark}}, \bibinfo {author} {\bibfnamefont {A.}~\bibnamefont {Gaj}}, \bibinfo
  {author} {\bibfnamefont {L.}~\bibnamefont {Feng}}, \ and\ \bibinfo {author}
  {\bibfnamefont {C.}~\bibnamefont {Chin}},\ }\href {\doibase
  10.1038/nature24272} {\bibfield  {journal} {\bibinfo  {journal} {Nature
  (London)}\ }\textbf {\bibinfo {volume} {551}},\ \bibinfo {pages} {356}
  (\bibinfo {year} {2017})}\BibitemShut {NoStop}%
\bibitem [{\citenamefont {Fu}\ \emph {et~al.}(2018)\citenamefont {Fu},
  \citenamefont {Feng}, \citenamefont {Anderson}, \citenamefont {Clark},
  \citenamefont {Hu}, \citenamefont {Andrade}, \citenamefont {Chin},\ and\
  \citenamefont {Levin}}]{Fu2018}%
  \BibitemOpen
  \bibfield  {author} {\bibinfo {author} {\bibfnamefont {H.}~\bibnamefont
  {Fu}}, \bibinfo {author} {\bibfnamefont {L.}~\bibnamefont {Feng}}, \bibinfo
  {author} {\bibfnamefont {B.~M.}\ \bibnamefont {Anderson}}, \bibinfo {author}
  {\bibfnamefont {L.~W.}\ \bibnamefont {Clark}}, \bibinfo {author}
  {\bibfnamefont {J.}~\bibnamefont {Hu}}, \bibinfo {author} {\bibfnamefont
  {J.~W.}\ \bibnamefont {Andrade}}, \bibinfo {author} {\bibfnamefont
  {C.}~\bibnamefont {Chin}}, \ and\ \bibinfo {author} {\bibfnamefont
  {K.}~\bibnamefont {Levin}},\ }\href {\doibase 10.1103/PhysRevLett.121.243001}
  {\bibfield  {journal} {\bibinfo  {journal} {Phys. Rev. Lett.}\ }\textbf
  {\bibinfo {volume} {121}},\ \bibinfo {pages} {243001} (\bibinfo {year}
  {2018})}\BibitemShut {NoStop}%
\bibitem [{\citenamefont {Chikkatur}\ \emph {et~al.}(2000)\citenamefont
  {Chikkatur}, \citenamefont {G\"orlitz}, \citenamefont {Stamper-Kurn},
  \citenamefont {Inouye}, \citenamefont {Gupta},\ and\ \citenamefont
  {Ketterle}}]{Chikkatur2000}%
  \BibitemOpen
  \bibfield  {author} {\bibinfo {author} {\bibfnamefont {A.~P.}\ \bibnamefont
  {Chikkatur}}, \bibinfo {author} {\bibfnamefont {A.}~\bibnamefont
  {G\"orlitz}}, \bibinfo {author} {\bibfnamefont {D.~M.}\ \bibnamefont
  {Stamper-Kurn}}, \bibinfo {author} {\bibfnamefont {S.}~\bibnamefont
  {Inouye}}, \bibinfo {author} {\bibfnamefont {S.}~\bibnamefont {Gupta}}, \
  and\ \bibinfo {author} {\bibfnamefont {W.}~\bibnamefont {Ketterle}},\ }\href
  {\doibase 10.1103/PhysRevLett.85.483} {\bibfield  {journal} {\bibinfo
  {journal} {Phys. Rev. Lett.}\ }\textbf {\bibinfo {volume} {85}},\ \bibinfo
  {pages} {483} (\bibinfo {year} {2000})}\BibitemShut {NoStop}%
\bibitem [{\citenamefont {Feng}\ \emph {et~al.}(2019)\citenamefont {Feng},
  \citenamefont {Hu}, \citenamefont {Clark},\ and\ \citenamefont
  {Chin}}]{Feng2019}%
  \BibitemOpen
  \bibfield  {author} {\bibinfo {author} {\bibfnamefont {L.}~\bibnamefont
  {Feng}}, \bibinfo {author} {\bibfnamefont {J.}~\bibnamefont {Hu}}, \bibinfo
  {author} {\bibfnamefont {L.~W.}\ \bibnamefont {Clark}}, \ and\ \bibinfo
  {author} {\bibfnamefont {C.}~\bibnamefont {Chin}},\ }\href {\doibase
  10.1126/science.aat5008} {\bibfield  {journal} {\bibinfo  {journal}
  {Science}\ }\textbf {\bibinfo {volume} {363}},\ \bibinfo {pages} {521}
  (\bibinfo {year} {2019})}\BibitemShut {NoStop}%
\bibitem [{\citenamefont {Hu}\ \emph {et~al.}(2019)\citenamefont {Hu},
  \citenamefont {Feng}, \citenamefont {Zhang},\ and\ \citenamefont
  {Chin}}]{Hu2019}%
  \BibitemOpen
  \bibfield  {author} {\bibinfo {author} {\bibfnamefont {J.}~\bibnamefont
  {Hu}}, \bibinfo {author} {\bibfnamefont {L.}~\bibnamefont {Feng}}, \bibinfo
  {author} {\bibfnamefont {Z.}~\bibnamefont {Zhang}}, \ and\ \bibinfo {author}
  {\bibfnamefont {C.}~\bibnamefont {Chin}},\ }\href {\doibase
  10.1038/s41567-019-0537-1} {\bibfield  {journal} {\bibinfo  {journal} {Nat.
  Phys.}\ }\textbf {\bibinfo {volume} {15}},\ \bibinfo {pages} {785} (\bibinfo
  {year} {2019})}\BibitemShut {NoStop}%
\bibitem [{\citenamefont {Krinner}\ \emph {et~al.}(2018)\citenamefont
  {Krinner}, \citenamefont {Stewart}, \citenamefont {Pazmi{\~{n}}o},
  \citenamefont {Kwon},\ and\ \citenamefont {Schneble}}]{Krinner2018}%
  \BibitemOpen
  \bibfield  {author} {\bibinfo {author} {\bibfnamefont {L.}~\bibnamefont
  {Krinner}}, \bibinfo {author} {\bibfnamefont {M.}~\bibnamefont {Stewart}},
  \bibinfo {author} {\bibfnamefont {A.}~\bibnamefont {Pazmi{\~{n}}o}}, \bibinfo
  {author} {\bibfnamefont {J.}~\bibnamefont {Kwon}}, \ and\ \bibinfo {author}
  {\bibfnamefont {D.}~\bibnamefont {Schneble}},\ }\href {\doibase
  10.1038/s41586-018-0348-z} {\bibfield  {journal} {\bibinfo  {journal} {Nature
  (London)}\ }\textbf {\bibinfo {volume} {559}},\ \bibinfo {pages} {589}
  (\bibinfo {year} {2018})}\BibitemShut {NoStop}%
\bibitem [{\citenamefont {Gross}\ \emph {et~al.}(2011)\citenamefont {Gross},
  \citenamefont {Strobel}, \citenamefont {Nicklas}, \citenamefont {Zibold},
  \citenamefont {Bar-Gill}, \citenamefont {Kurizki},\ and\ \citenamefont
  {Oberthaler}}]{Gross2011}%
  \BibitemOpen
  \bibfield  {author} {\bibinfo {author} {\bibfnamefont {C.}~\bibnamefont
  {Gross}}, \bibinfo {author} {\bibfnamefont {H.}~\bibnamefont {Strobel}},
  \bibinfo {author} {\bibfnamefont {E.}~\bibnamefont {Nicklas}}, \bibinfo
  {author} {\bibfnamefont {T.}~\bibnamefont {Zibold}}, \bibinfo {author}
  {\bibfnamefont {N.}~\bibnamefont {Bar-Gill}}, \bibinfo {author}
  {\bibfnamefont {G.}~\bibnamefont {Kurizki}}, \ and\ \bibinfo {author}
  {\bibfnamefont {M.~K.}\ \bibnamefont {Oberthaler}},\ }\href {\doibase
  10.1038/nature10654} {\bibfield  {journal} {\bibinfo  {journal} {Nature
  (London)}\ }\textbf {\bibinfo {volume} {480}},\ \bibinfo {pages} {219}
  (\bibinfo {year} {2011})}\BibitemShut {NoStop}%
\bibitem [{\citenamefont {L{\"{u}}cke}\ \emph {et~al.}(2011)\citenamefont
  {L{\"{u}}cke}, \citenamefont {Scherer}, \citenamefont {Kruse}, \citenamefont
  {Pezz{\'{e}}}, \citenamefont {Deuretzbacher}, \citenamefont {Hyllus},
  \citenamefont {Topic}, \citenamefont {Peise}, \citenamefont {Ertmer},
  \citenamefont {Arlt}, \citenamefont {Santos}, \citenamefont {Smerzi},\ and\
  \citenamefont {Klempt}}]{Lucke2011}%
  \BibitemOpen
  \bibfield  {author} {\bibinfo {author} {\bibfnamefont {B.}~\bibnamefont
  {L{\"{u}}cke}}, \bibinfo {author} {\bibfnamefont {M.}~\bibnamefont
  {Scherer}}, \bibinfo {author} {\bibfnamefont {J.}~\bibnamefont {Kruse}},
  \bibinfo {author} {\bibfnamefont {L.}~\bibnamefont {Pezz{\'{e}}}}, \bibinfo
  {author} {\bibfnamefont {F.}~\bibnamefont {Deuretzbacher}}, \bibinfo {author}
  {\bibfnamefont {P.}~\bibnamefont {Hyllus}}, \bibinfo {author} {\bibfnamefont
  {O.}~\bibnamefont {Topic}}, \bibinfo {author} {\bibfnamefont
  {J.}~\bibnamefont {Peise}}, \bibinfo {author} {\bibfnamefont
  {W.}~\bibnamefont {Ertmer}}, \bibinfo {author} {\bibfnamefont
  {J.}~\bibnamefont {Arlt}}, \bibinfo {author} {\bibfnamefont {L.}~\bibnamefont
  {Santos}}, \bibinfo {author} {\bibfnamefont {A.}~\bibnamefont {Smerzi}}, \
  and\ \bibinfo {author} {\bibfnamefont {C.}~\bibnamefont {Klempt}},\ }\href
  {\doibase 10.1126/science.1208798} {\bibfield  {journal} {\bibinfo  {journal}
  {Science}\ }\textbf {\bibinfo {volume} {334}},\ \bibinfo {pages} {773}
  (\bibinfo {year} {2011})}\BibitemShut {NoStop}%
\bibitem [{\citenamefont {Hamley}\ \emph {et~al.}(2012)\citenamefont {Hamley},
  \citenamefont {Gerving}, \citenamefont {Hoang}, \citenamefont {Bookjans},\
  and\ \citenamefont {Chapman}}]{Hamley2012}%
  \BibitemOpen
  \bibfield  {author} {\bibinfo {author} {\bibfnamefont {C.~D.}\ \bibnamefont
  {Hamley}}, \bibinfo {author} {\bibfnamefont {C.~S.}\ \bibnamefont {Gerving}},
  \bibinfo {author} {\bibfnamefont {T.~M.}\ \bibnamefont {Hoang}}, \bibinfo
  {author} {\bibfnamefont {E.~M.}\ \bibnamefont {Bookjans}}, \ and\ \bibinfo
  {author} {\bibfnamefont {M.~S.}\ \bibnamefont {Chapman}},\ }\href
  {https://doi.org/10.1038/nphys2245} {\bibfield  {journal} {\bibinfo
  {journal} {Nat. Phys.}\ }\textbf {\bibinfo {volume} {8}},\ \bibinfo {pages}
  {305} (\bibinfo {year} {2012})}\BibitemShut {NoStop}%
\bibitem [{\citenamefont {Lange}\ \emph {et~al.}(2018)\citenamefont {Lange},
  \citenamefont {Peise}, \citenamefont {L{\"{u}}cke}, \citenamefont {Kruse},
  \citenamefont {Vitagliano}, \citenamefont {Apellaniz}, \citenamefont
  {Kleinmann}, \citenamefont {T{\'{o}}th},\ and\ \citenamefont
  {Klempt}}]{Lange2018}%
  \BibitemOpen
  \bibfield  {author} {\bibinfo {author} {\bibfnamefont {K.}~\bibnamefont
  {Lange}}, \bibinfo {author} {\bibfnamefont {J.}~\bibnamefont {Peise}},
  \bibinfo {author} {\bibfnamefont {B.}~\bibnamefont {L{\"{u}}cke}}, \bibinfo
  {author} {\bibfnamefont {I.}~\bibnamefont {Kruse}}, \bibinfo {author}
  {\bibfnamefont {G.}~\bibnamefont {Vitagliano}}, \bibinfo {author}
  {\bibfnamefont {I.}~\bibnamefont {Apellaniz}}, \bibinfo {author}
  {\bibfnamefont {M.}~\bibnamefont {Kleinmann}}, \bibinfo {author}
  {\bibfnamefont {G.}~\bibnamefont {T{\'{o}}th}}, \ and\ \bibinfo {author}
  {\bibfnamefont {C.}~\bibnamefont {Klempt}},\ }\href {\doibase
  10.1126/science.aao2035} {\bibfield  {journal} {\bibinfo  {journal}
  {Science}\ }\textbf {\bibinfo {volume} {360}},\ \bibinfo {pages} {416}
  (\bibinfo {year} {2018})}\BibitemShut {NoStop}%
\bibitem [{\citenamefont {Kunkel}\ \emph {et~al.}(2018)\citenamefont {Kunkel},
  \citenamefont {Pr{\"{u}}fer}, \citenamefont {Strobel}, \citenamefont
  {Linnemann}, \citenamefont {Fr{\"{o}}lian}, \citenamefont {Gasenzer},
  \citenamefont {G{\"{a}}rttner},\ and\ \citenamefont
  {Oberthaler}}]{Kunkel2018}%
  \BibitemOpen
  \bibfield  {author} {\bibinfo {author} {\bibfnamefont {P.}~\bibnamefont
  {Kunkel}}, \bibinfo {author} {\bibfnamefont {M.}~\bibnamefont
  {Pr{\"{u}}fer}}, \bibinfo {author} {\bibfnamefont {H.}~\bibnamefont
  {Strobel}}, \bibinfo {author} {\bibfnamefont {D.}~\bibnamefont {Linnemann}},
  \bibinfo {author} {\bibfnamefont {A.}~\bibnamefont {Fr{\"{o}}lian}}, \bibinfo
  {author} {\bibfnamefont {T.}~\bibnamefont {Gasenzer}}, \bibinfo {author}
  {\bibfnamefont {M.}~\bibnamefont {G{\"{a}}rttner}}, \ and\ \bibinfo {author}
  {\bibfnamefont {M.~K.}\ \bibnamefont {Oberthaler}},\ }\href {\doibase
  10.1126/science.aao2254} {\bibfield  {journal} {\bibinfo  {journal}
  {Science}\ }\textbf {\bibinfo {volume} {360}},\ \bibinfo {pages} {413}
  (\bibinfo {year} {2018})}\BibitemShut {NoStop}%
\bibitem [{\citenamefont {Muessel}\ \emph {et~al.}(2014)\citenamefont
  {Muessel}, \citenamefont {Strobel}, \citenamefont {Linnemann}, \citenamefont
  {Hume},\ and\ \citenamefont {Oberthaler}}]{Muessel2014}%
  \BibitemOpen
  \bibfield  {author} {\bibinfo {author} {\bibfnamefont {W.}~\bibnamefont
  {Muessel}}, \bibinfo {author} {\bibfnamefont {H.}~\bibnamefont {Strobel}},
  \bibinfo {author} {\bibfnamefont {D.}~\bibnamefont {Linnemann}}, \bibinfo
  {author} {\bibfnamefont {D.~B.}\ \bibnamefont {Hume}}, \ and\ \bibinfo
  {author} {\bibfnamefont {M.~K.}\ \bibnamefont {Oberthaler}},\ }\href
  {\doibase 10.1103/PhysRevLett.113.103004} {\bibfield  {journal} {\bibinfo
  {journal} {Phys. Rev. Lett.}\ }\textbf {\bibinfo {volume} {113}},\ \bibinfo
  {pages} {103004} (\bibinfo {year} {2014})}\BibitemShut {NoStop}%
\bibitem [{\citenamefont {Kruse}\ \emph {et~al.}(2016)\citenamefont {Kruse},
  \citenamefont {Lange}, \citenamefont {Peise}, \citenamefont {L\"ucke},
  \citenamefont {Pezz\`e}, \citenamefont {Arlt}, \citenamefont {Ertmer},
  \citenamefont {Lisdat}, \citenamefont {Santos}, \citenamefont {Smerzi},\ and\
  \citenamefont {Klempt}}]{Kruse2016}%
  \BibitemOpen
  \bibfield  {author} {\bibinfo {author} {\bibfnamefont {I.}~\bibnamefont
  {Kruse}}, \bibinfo {author} {\bibfnamefont {K.}~\bibnamefont {Lange}},
  \bibinfo {author} {\bibfnamefont {J.}~\bibnamefont {Peise}}, \bibinfo
  {author} {\bibfnamefont {B.}~\bibnamefont {L\"ucke}}, \bibinfo {author}
  {\bibfnamefont {L.}~\bibnamefont {Pezz\`e}}, \bibinfo {author} {\bibfnamefont
  {J.}~\bibnamefont {Arlt}}, \bibinfo {author} {\bibfnamefont {W.}~\bibnamefont
  {Ertmer}}, \bibinfo {author} {\bibfnamefont {C.}~\bibnamefont {Lisdat}},
  \bibinfo {author} {\bibfnamefont {L.}~\bibnamefont {Santos}}, \bibinfo
  {author} {\bibfnamefont {A.}~\bibnamefont {Smerzi}}, \ and\ \bibinfo {author}
  {\bibfnamefont {C.}~\bibnamefont {Klempt}},\ }\href {\doibase
  10.1103/PhysRevLett.117.143004} {\bibfield  {journal} {\bibinfo  {journal}
  {Phys. Rev. Lett.}\ }\textbf {\bibinfo {volume} {117}},\ \bibinfo {pages}
  {143004} (\bibinfo {year} {2016})}\BibitemShut {NoStop}%
\bibitem [{\citenamefont {Pezz\`e}\ \emph {et~al.}(2018)\citenamefont
  {Pezz\`e}, \citenamefont {Smerzi}, \citenamefont {Oberthaler}, \citenamefont
  {Schmied},\ and\ \citenamefont {Treutlein}}]{Pezze2018}%
  \BibitemOpen
  \bibfield  {author} {\bibinfo {author} {\bibfnamefont {L.}~\bibnamefont
  {Pezz\`e}}, \bibinfo {author} {\bibfnamefont {A.}~\bibnamefont {Smerzi}},
  \bibinfo {author} {\bibfnamefont {M.~K.}\ \bibnamefont {Oberthaler}},
  \bibinfo {author} {\bibfnamefont {R.}~\bibnamefont {Schmied}}, \ and\
  \bibinfo {author} {\bibfnamefont {P.}~\bibnamefont {Treutlein}},\ }\href
  {\doibase 10.1103/RevModPhys.90.035005} {\bibfield  {journal} {\bibinfo
  {journal} {Rev. Mod. Phys.}\ }\textbf {\bibinfo {volume} {90}},\ \bibinfo
  {pages} {035005} (\bibinfo {year} {2018})}\BibitemShut {NoStop}%
\bibitem [{\citenamefont {Einstein}\ \emph {et~al.}(1935)\citenamefont
  {Einstein}, \citenamefont {Podolsky},\ and\ \citenamefont
  {Rosen}}]{Einstein1935}%
  \BibitemOpen
  \bibfield  {author} {\bibinfo {author} {\bibfnamefont {A.}~\bibnamefont
  {Einstein}}, \bibinfo {author} {\bibfnamefont {B.}~\bibnamefont {Podolsky}},
  \ and\ \bibinfo {author} {\bibfnamefont {N.}~\bibnamefont {Rosen}},\ }\href
  {\doibase 10.1103/PhysRev.47.777} {\bibfield  {journal} {\bibinfo  {journal}
  {Phys. Rev.}\ }\textbf {\bibinfo {volume} {47}},\ \bibinfo {pages} {777}
  (\bibinfo {year} {1935})}\BibitemShut {NoStop}%
\bibitem [{\citenamefont {Pu}\ and\ \citenamefont {Meystre}(2000)}]{Pu2000}%
  \BibitemOpen
  \bibfield  {author} {\bibinfo {author} {\bibfnamefont {H.}~\bibnamefont
  {Pu}}\ and\ \bibinfo {author} {\bibfnamefont {P.}~\bibnamefont {Meystre}},\
  }\href {\doibase 10.1103/PhysRevLett.85.3987} {\bibfield  {journal} {\bibinfo
   {journal} {Phys. Rev. Lett.}\ }\textbf {\bibinfo {volume} {85}},\ \bibinfo
  {pages} {3987} (\bibinfo {year} {2000})}\BibitemShut {NoStop}%
\bibitem [{\citenamefont {Duan}\ \emph {et~al.}(2000)\citenamefont {Duan},
  \citenamefont {S\o{}rensen}, \citenamefont {Cirac},\ and\ \citenamefont
  {Zoller}}]{Duan2000}%
  \BibitemOpen
  \bibfield  {author} {\bibinfo {author} {\bibfnamefont {L.-M.}\ \bibnamefont
  {Duan}}, \bibinfo {author} {\bibfnamefont {A.}~\bibnamefont {S\o{}rensen}},
  \bibinfo {author} {\bibfnamefont {J.~I.}\ \bibnamefont {Cirac}}, \ and\
  \bibinfo {author} {\bibfnamefont {P.}~\bibnamefont {Zoller}},\ }\href
  {\doibase 10.1103/PhysRevLett.85.3991} {\bibfield  {journal} {\bibinfo
  {journal} {Phys. Rev. Lett.}\ }\textbf {\bibinfo {volume} {85}},\ \bibinfo
  {pages} {3991} (\bibinfo {year} {2000})}\BibitemShut {NoStop}%
\bibitem [{\citenamefont {Huh}\ \emph {et~al.}(2020)\citenamefont {Huh},
  \citenamefont {Kim}, \citenamefont {Kwon},\ and\ \citenamefont
  {Choi}}]{Huh2020}%
  \BibitemOpen
  \bibfield  {author} {\bibinfo {author} {\bibfnamefont {S.}~\bibnamefont
  {Huh}}, \bibinfo {author} {\bibfnamefont {K.}~\bibnamefont {Kim}}, \bibinfo
  {author} {\bibfnamefont {K.}~\bibnamefont {Kwon}}, \ and\ \bibinfo {author}
  {\bibfnamefont {J.-y.}\ \bibnamefont {Choi}},\ }\href {\doibase
  10.1103/PhysRevResearch.2.033471} {\bibfield  {journal} {\bibinfo  {journal}
  {Phys. Rev. Research}\ }\textbf {\bibinfo {volume} {2}},\ \bibinfo {pages}
  {033471} (\bibinfo {year} {2020})}\BibitemShut {NoStop}%
\bibitem [{\citenamefont {Kawaguchi}\ and\ \citenamefont
  {Ueda}(2012)}]{Kawaguchi2012}%
  \BibitemOpen
  \bibfield  {author} {\bibinfo {author} {\bibfnamefont {Y.}~\bibnamefont
  {Kawaguchi}}\ and\ \bibinfo {author} {\bibfnamefont {M.}~\bibnamefont
  {Ueda}},\ }\href {\doibase https://doi.org/10.1016/j.physrep.2012.07.005}
  {\bibfield  {journal} {\bibinfo  {journal} {Physics Reports}\ }\textbf
  {\bibinfo {volume} {520}},\ \bibinfo {pages} {253 } (\bibinfo {year}
  {2012})}\BibitemShut {NoStop}%
\bibitem [{\citenamefont {Gerbier}\ \emph {et~al.}(2006)\citenamefont
  {Gerbier}, \citenamefont {Widera}, \citenamefont {F\"olling}, \citenamefont
  {Mandel},\ and\ \citenamefont {Bloch}}]{Gerbier2006}%
  \BibitemOpen
  \bibfield  {author} {\bibinfo {author} {\bibfnamefont {F.}~\bibnamefont
  {Gerbier}}, \bibinfo {author} {\bibfnamefont {A.}~\bibnamefont {Widera}},
  \bibinfo {author} {\bibfnamefont {S.}~\bibnamefont {F\"olling}}, \bibinfo
  {author} {\bibfnamefont {O.}~\bibnamefont {Mandel}}, \ and\ \bibinfo {author}
  {\bibfnamefont {I.}~\bibnamefont {Bloch}},\ }\href {\doibase
  10.1103/PhysRevA.73.041602} {\bibfield  {journal} {\bibinfo  {journal} {Phys.
  Rev. A}\ }\textbf {\bibinfo {volume} {73}},\ \bibinfo {pages} {041602}
  (\bibinfo {year} {2006})}\BibitemShut {NoStop}%
\bibitem [{\citenamefont {SM}(2021)}]{SM2021}%
  \BibitemOpen
 {See Supplement Material.}%
\bibitem [{\citenamefont {Vogels}\ \emph {et~al.}(2002)\citenamefont {Vogels},
  \citenamefont {Xu},\ and\ \citenamefont {Ketterle}}]{Vogels2002}%
  \BibitemOpen
  \bibfield  {author} {\bibinfo {author} {\bibfnamefont {J.~M.}\ \bibnamefont
  {Vogels}}, \bibinfo {author} {\bibfnamefont {K.}~\bibnamefont {Xu}}, \ and\
  \bibinfo {author} {\bibfnamefont {W.}~\bibnamefont {Ketterle}},\ }\href
  {\doibase 10.1103/PhysRevLett.89.020401} {\bibfield  {journal} {\bibinfo
  {journal} {Phys. Rev. Lett.}\ }\textbf {\bibinfo {volume} {89}},\ \bibinfo
  {pages} {020401} (\bibinfo {year} {2002})}\BibitemShut {NoStop}%
\bibitem [{\citenamefont {Klempt}\ \emph {et~al.}(2010)\citenamefont {Klempt},
  \citenamefont {Topic}, \citenamefont {Gebreyesus}, \citenamefont {Scherer},
  \citenamefont {Henninger}, \citenamefont {Hyllus}, \citenamefont {Ertmer},
  \citenamefont {Santos},\ and\ \citenamefont {Arlt}}]{Klempt2010}%
  \BibitemOpen
  \bibfield  {author} {\bibinfo {author} {\bibfnamefont {C.}~\bibnamefont
  {Klempt}}, \bibinfo {author} {\bibfnamefont {O.}~\bibnamefont {Topic}},
  \bibinfo {author} {\bibfnamefont {G.}~\bibnamefont {Gebreyesus}}, \bibinfo
  {author} {\bibfnamefont {M.}~\bibnamefont {Scherer}}, \bibinfo {author}
  {\bibfnamefont {T.}~\bibnamefont {Henninger}}, \bibinfo {author}
  {\bibfnamefont {P.}~\bibnamefont {Hyllus}}, \bibinfo {author} {\bibfnamefont
  {W.}~\bibnamefont {Ertmer}}, \bibinfo {author} {\bibfnamefont
  {L.}~\bibnamefont {Santos}}, \ and\ \bibinfo {author} {\bibfnamefont {J.~J.}\
  \bibnamefont {Arlt}},\ }\href {\doibase 10.1103/PhysRevLett.104.195303}
  {\bibfield  {journal} {\bibinfo  {journal} {Phys. Rev. Lett.}\ }\textbf
  {\bibinfo {volume} {104}},\ \bibinfo {pages} {195303} (\bibinfo {year}
  {2010})}\BibitemShut {NoStop}%
\bibitem [{\citenamefont {De}\ \emph {et~al.}(2014)\citenamefont {De},
  \citenamefont {Campbell}, \citenamefont {Price}, \citenamefont {Putra},
  \citenamefont {Anderson},\ and\ \citenamefont {Spielman}}]{De2014}%
  \BibitemOpen
  \bibfield  {author} {\bibinfo {author} {\bibfnamefont {S.}~\bibnamefont
  {De}}, \bibinfo {author} {\bibfnamefont {D.~L.}\ \bibnamefont {Campbell}},
  \bibinfo {author} {\bibfnamefont {R.~M.}\ \bibnamefont {Price}}, \bibinfo
  {author} {\bibfnamefont {A.}~\bibnamefont {Putra}}, \bibinfo {author}
  {\bibfnamefont {B.~M.}\ \bibnamefont {Anderson}}, \ and\ \bibinfo {author}
  {\bibfnamefont {I.~B.}\ \bibnamefont {Spielman}},\ }\href {\doibase
  10.1103/PhysRevA.89.033631} {\bibfield  {journal} {\bibinfo  {journal} {Phys.
  Rev. A}\ }\textbf {\bibinfo {volume} {89}},\ \bibinfo {pages} {033631}
  (\bibinfo {year} {2014})}\BibitemShut {NoStop}%
\bibitem [{\citenamefont {Hanbury~Brown}\ and\ \citenamefont
  {Twiss}(1956)}]{Hanbury1956}%
  \BibitemOpen
  \bibfield  {author} {\bibinfo {author} {\bibfnamefont {R.}~\bibnamefont
  {Hanbury~Brown}}\ and\ \bibinfo {author} {\bibfnamefont {R.~Q.}\ \bibnamefont
  {Twiss}},\ }\href {\doibase 10.1038/177027a0} {\bibfield  {journal} {\bibinfo
   {journal} {Nature (London)}\ }\textbf {\bibinfo {volume} {177}},\ \bibinfo
  {pages} {27} (\bibinfo {year} {1956})}\BibitemShut {NoStop}%
\bibitem [{\citenamefont {Wu}\ and\ \citenamefont {Zhai}(2019)}]{Wu2019}%
  \BibitemOpen
  \bibfield  {author} {\bibinfo {author} {\bibfnamefont {Z.}~\bibnamefont
  {Wu}}\ and\ \bibinfo {author} {\bibfnamefont {H.}~\bibnamefont {Zhai}},\
  }\href {\doibase 10.1103/PhysRevA.99.063624} {\bibfield  {journal} {\bibinfo
  {journal} {Phys. Rev. A}\ }\textbf {\bibinfo {volume} {99}},\ \bibinfo
  {pages} {063624} (\bibinfo {year} {2019})}\BibitemShut {NoStop}%
\bibitem [{\citenamefont {Mias}\ \emph {et~al.}(2008)\citenamefont {Mias},
  \citenamefont {Cooper},\ and\ \citenamefont {Girvin}}]{Mias2008}%
  \BibitemOpen
  \bibfield  {author} {\bibinfo {author} {\bibfnamefont {G.~I.}\ \bibnamefont
  {Mias}}, \bibinfo {author} {\bibfnamefont {N.~R.}\ \bibnamefont {Cooper}}, \
  and\ \bibinfo {author} {\bibfnamefont {S.~M.}\ \bibnamefont {Girvin}},\
  }\href {\doibase 10.1103/PhysRevA.77.023616} {\bibfield  {journal} {\bibinfo
  {journal} {Phys. Rev. A}\ }\textbf {\bibinfo {volume} {77}},\ \bibinfo
  {pages} {023616} (\bibinfo {year} {2008})}\BibitemShut {NoStop}%
\bibitem [{\citenamefont {Iskhakov}\ \emph {et~al.}(2012)\citenamefont
  {Iskhakov}, \citenamefont {Agafonov}, \citenamefont {Chekhova},\ and\
  \citenamefont {Leuchs}}]{Iskhakov2012}%
  \BibitemOpen
  \bibfield  {author} {\bibinfo {author} {\bibfnamefont {T.~S.}\ \bibnamefont
  {Iskhakov}}, \bibinfo {author} {\bibfnamefont {I.~N.}\ \bibnamefont
  {Agafonov}}, \bibinfo {author} {\bibfnamefont {M.~V.}\ \bibnamefont
  {Chekhova}}, \ and\ \bibinfo {author} {\bibfnamefont {G.}~\bibnamefont
  {Leuchs}},\ }\href {\doibase 10.1103/PhysRevLett.109.150502} {\bibfield
  {journal} {\bibinfo  {journal} {Phys. Rev. Lett.}\ }\textbf {\bibinfo
  {volume} {109}},\ \bibinfo {pages} {150502} (\bibinfo {year}
  {2012})}\BibitemShut {NoStop}%
\bibitem [{\citenamefont {Me\ifmmode \check{z}\else \v{z}\fi{}nar\ifmmode
  \check{s}\else \v{s}\fi{}i\ifmmode~\check{c}\else \v{c}\fi{}}\ \emph
  {et~al.}(2020)\citenamefont {Me\ifmmode \check{z}\else \v{z}\fi{}nar\ifmmode
  \check{s}\else \v{s}\fi{}i\ifmmode~\check{c}\else \v{c}\fi{}}, \citenamefont
  {\ifmmode~\check{Z}\else \v{Z}\fi{}itko}, \citenamefont {Arh}, \citenamefont
  {Gosar}, \citenamefont {Zupani\ifmmode~\check{c}\else \v{c}\fi{}},\ and\
  \citenamefont {Jegli\ifmmode~\check{c}\else \v{c}\fi{}}}]{Meznarsic2020}%
  \BibitemOpen
  \bibfield  {author} {\bibinfo {author} {\bibfnamefont {T.}~\bibnamefont
  {Me\ifmmode \check{z}\else \v{z}\fi{}nar\ifmmode \check{s}\else
  \v{s}\fi{}i\ifmmode~\check{c}\else \v{c}\fi{}}}, \bibinfo {author}
  {\bibfnamefont {R.}~\bibnamefont {\ifmmode~\check{Z}\else \v{Z}\fi{}itko}},
  \bibinfo {author} {\bibfnamefont {T.}~\bibnamefont {Arh}}, \bibinfo {author}
  {\bibfnamefont {K.}~\bibnamefont {Gosar}}, \bibinfo {author} {\bibfnamefont
  {E.}~\bibnamefont {Zupani\ifmmode~\check{c}\else \v{c}\fi{}}}, \ and\
  \bibinfo {author} {\bibfnamefont {P.}~\bibnamefont
  {Jegli\ifmmode~\check{c}\else \v{c}\fi{}}},\ }\href {\doibase
  10.1103/PhysRevA.101.031601} {\bibfield  {journal} {\bibinfo  {journal}
  {Phys. Rev. A}\ }\textbf {\bibinfo {volume} {101}},\ \bibinfo {pages}
  {031601(R)} (\bibinfo {year} {2020})}\BibitemShut {NoStop}%
\bibitem [{\citenamefont {Bell}(1964)}]{Bell1964}%
  \BibitemOpen
  \bibfield  {author} {\bibinfo {author} {\bibfnamefont {J.~S.}\ \bibnamefont
  {Bell}},\ }\href {\doibase 10.1103/PhysicsPhysiqueFizika.1.195} {\bibfield
  {journal} {\bibinfo  {journal} {Physics Physique Fizika}\ }\textbf {\bibinfo
  {volume} {1}},\ \bibinfo {pages} {195} (\bibinfo {year} {1964})}\BibitemShut
  {NoStop}%
\bibitem [{\citenamefont {Mathy}\ \emph {et~al.}(2012)\citenamefont {Mathy},
  \citenamefont {Zvonarev},\ and\ \citenamefont {Demler}}]{Mathy2012}%
  \BibitemOpen
  \bibfield  {author} {\bibinfo {author} {\bibfnamefont {C.}~\bibnamefont
  {Mathy}}, \bibinfo {author} {\bibfnamefont {M.}~\bibnamefont {Zvonarev}}, \
  and\ \bibinfo {author} {\bibfnamefont {E.}~\bibnamefont {Demler}},\ }\href
  {\doibase 10.1038/nphys2455} {\bibfield  {journal} {\bibinfo  {journal} {Nat.
  Phys.}\ }\textbf {\bibinfo {volume} {8}},\ \bibinfo {pages} {881} (\bibinfo
  {year} {2012})}\BibitemShut {NoStop}%
\bibitem [{\citenamefont {Simon}\ and\ \citenamefont
  {Bouwmeester}(2003)}]{Simon2003}%
  \BibitemOpen
  \bibfield  {author} {\bibinfo {author} {\bibfnamefont {C.}~\bibnamefont
  {Simon}}\ and\ \bibinfo {author} {\bibfnamefont {D.}~\bibnamefont
  {Bouwmeester}},\ }\href {\doibase 10.1103/PhysRevLett.91.053601} {\bibfield
  {journal} {\bibinfo  {journal} {Phys. Rev. Lett.}\ }\textbf {\bibinfo
  {volume} {91}},\ \bibinfo {pages} {053601} (\bibinfo {year}
  {2003})}\BibitemShut {NoStop}%
\bibitem [{\citenamefont {Weiner}\ \emph {et~al.}(1999)\citenamefont {Weiner},
  \citenamefont {Bagnato}, \citenamefont {Zilio},\ and\ \citenamefont
  {Julienne}}]{Weiner1999}%
  \BibitemOpen
  \bibfield  {author} {\bibinfo {author} {\bibfnamefont {J.}~\bibnamefont
  {Weiner}}, \bibinfo {author} {\bibfnamefont {V.~S.}\ \bibnamefont {Bagnato}},
  \bibinfo {author} {\bibfnamefont {S.}~\bibnamefont {Zilio}}, \ and\ \bibinfo
  {author} {\bibfnamefont {P.~S.}\ \bibnamefont {Julienne}},\ }\href {\doibase
  10.1103/RevModPhys.71.1} {\bibfield  {journal} {\bibinfo  {journal} {Rev.
  Mod. Phys.}\ }\textbf {\bibinfo {volume} {71}},\ \bibinfo {pages} {1}
  (\bibinfo {year} {1999})}\BibitemShut {NoStop}%
\bibitem [{\citenamefont {Gerton}\ \emph {et~al.}(1999)\citenamefont {Gerton},
  \citenamefont {Sackett}, \citenamefont {Frew},\ and\ \citenamefont
  {Hulet}}]{Gerton1999}%
  \BibitemOpen
  \bibfield  {author} {\bibinfo {author} {\bibfnamefont {J.~M.}\ \bibnamefont
  {Gerton}}, \bibinfo {author} {\bibfnamefont {C.~A.}\ \bibnamefont {Sackett}},
  \bibinfo {author} {\bibfnamefont {B.~J.}\ \bibnamefont {Frew}}, \ and\
  \bibinfo {author} {\bibfnamefont {R.~G.}\ \bibnamefont {Hulet}},\ }\href
  {\doibase 10.1103/PhysRevA.59.1514} {\bibfield  {journal} {\bibinfo
  {journal} {Phys. Rev. A}\ }\textbf {\bibinfo {volume} {59}},\ \bibinfo
  {pages} {1514} (\bibinfo {year} {1999})}\BibitemShut {NoStop}%
\bibitem [{\citenamefont {Itano}\ \emph {et~al.}(1993)\citenamefont {Itano},
  \citenamefont {Bergquist}, \citenamefont {Bollinger}, \citenamefont
  {Gilligan}, \citenamefont {Heinzen}, \citenamefont {Moore}, \citenamefont
  {Raizen},\ and\ \citenamefont {Wineland}}]{Itano1993}%
  \BibitemOpen
  \bibfield  {author} {\bibinfo {author} {\bibfnamefont {W.~M.}\ \bibnamefont
  {Itano}}, \bibinfo {author} {\bibfnamefont {J.~C.}\ \bibnamefont
  {Bergquist}}, \bibinfo {author} {\bibfnamefont {J.~J.}\ \bibnamefont
  {Bollinger}}, \bibinfo {author} {\bibfnamefont {J.~M.}\ \bibnamefont
  {Gilligan}}, \bibinfo {author} {\bibfnamefont {D.~J.}\ \bibnamefont
  {Heinzen}}, \bibinfo {author} {\bibfnamefont {F.~L.}\ \bibnamefont {Moore}},
  \bibinfo {author} {\bibfnamefont {M.~G.}\ \bibnamefont {Raizen}}, \ and\
  \bibinfo {author} {\bibfnamefont {D.~J.}\ \bibnamefont {Wineland}},\ }\href
  {\doibase 10.1103/PhysRevA.47.3554} {\bibfield  {journal} {\bibinfo
  {journal} {Phys. Rev. A}\ }\textbf {\bibinfo {volume} {47}},\ \bibinfo
  {pages} {3554} (\bibinfo {year} {1993})}\BibitemShut {NoStop}%
\bibitem [{\citenamefont {Est\`{e}ve}\ \emph {et~al.}(2007)\citenamefont
  {Est\`{e}ve}, \citenamefont {Gross}, \citenamefont {A.}, \citenamefont {S.},\
  and\ \citenamefont {Oberthaler}}]{Esteve2008}%
  \BibitemOpen
  \bibfield  {author} {\bibinfo {author} {\bibfnamefont {J.}~\bibnamefont
  {Est\`{e}ve}}, \bibinfo {author} {\bibfnamefont {C.}~\bibnamefont {Gross}},
  \bibinfo {author} {\bibfnamefont {W.}~\bibnamefont {A.}}, \bibinfo {author}
  {\bibfnamefont {G.}~\bibnamefont {S.}}, \ and\ \bibinfo {author}
  {\bibfnamefont {M.~K.}\ \bibnamefont {Oberthaler}},\ }\href {\doibase
  10.1038/nature07332} {\bibfield  {journal} {\bibinfo  {journal} {Nature
  (London)}\ }\textbf {\bibinfo {volume} {455}},\ \bibinfo {pages} {1216}
  (\bibinfo {year} {2007})}\BibitemShut {NoStop}%
\bibitem [{\citenamefont {Reidel}\ \emph {et~al.}(2010)\citenamefont {Reidel},
  \citenamefont {B\"{o}hi}, \citenamefont {Li}, \citenamefont {H\"{a}nsch},
  \citenamefont {Sinatra},\ and\ \citenamefont {Treutlein}}]{Reidel2010}%
  \BibitemOpen
  \bibfield  {author} {\bibinfo {author} {\bibfnamefont {M.~F.}\ \bibnamefont
  {Reidel}}, \bibinfo {author} {\bibfnamefont {P.}~\bibnamefont {B\"{o}hi}},
  \bibinfo {author} {\bibfnamefont {Y.}~\bibnamefont {Li}}, \bibinfo {author}
  {\bibfnamefont {T.~W.}\ \bibnamefont {H\"{a}nsch}}, \bibinfo {author}
  {\bibfnamefont {A.}~\bibnamefont {Sinatra}}, \ and\ \bibinfo {author}
  {\bibfnamefont {P.}~\bibnamefont {Treutlein}},\ }\href {\doibase
  10.1038/nature08988} {\bibfield  {journal} {\bibinfo  {journal} {Nature
  (London)}\ }\textbf {\bibinfo {volume} {464}},\ \bibinfo {pages} {1170}
  (\bibinfo {year} {2010})}\BibitemShut {NoStop}%
\bibitem [{\citenamefont {Muessel}\ \emph {et~al.}(2013)\citenamefont
  {Muessel}, \citenamefont {Strobel}, \citenamefont {Joos}, \citenamefont
  {Nicklas}, \citenamefont {Stroescu}, \citenamefont {Tomkovi\u{c}},
  \citenamefont {Hume},\ and\ \citenamefont {Oberthaler}}]{Muessel2013}%
  \BibitemOpen
  \bibfield  {author} {\bibinfo {author} {\bibfnamefont {W.}~\bibnamefont
  {Muessel}}, \bibinfo {author} {\bibfnamefont {H.}~\bibnamefont {Strobel}},
  \bibinfo {author} {\bibfnamefont {M.}~\bibnamefont {Joos}}, \bibinfo {author}
  {\bibfnamefont {E.}~\bibnamefont {Nicklas}}, \bibinfo {author} {\bibfnamefont
  {I.}~\bibnamefont {Stroescu}}, \bibinfo {author} {\bibfnamefont
  {J.}~\bibnamefont {Tomkovi\u{c}}}, \bibinfo {author} {\bibfnamefont {D.~H.}\
  \bibnamefont {Hume}}, \ and\ \bibinfo {author} {\bibfnamefont {M.~K.}\
  \bibnamefont {Oberthaler}},\ }\href {\doibase 10.1007/s00340-013-5553-8}
  {\bibfield  {journal} {\bibinfo  {journal} {Appl. Phys. B}\ }\textbf
  {\bibinfo {volume} {113}},\ \bibinfo {pages} {69} (\bibinfo {year}
  {2013})}\BibitemShut {NoStop}%
\bibitem [{\citenamefont {Hueck}\ \emph {et~al.}(2017)\citenamefont {Hueck},
  \citenamefont {Luick}, \citenamefont {Sobirey}, \citenamefont {Siegl},
  \citenamefont {Lompe}, \citenamefont {Moritz}, \citenamefont {Clark},\ and\
  \citenamefont {Chin}}]{Hueck2017}%
  \BibitemOpen
  \bibfield  {author} {\bibinfo {author} {\bibfnamefont {K.}~\bibnamefont
  {Hueck}}, \bibinfo {author} {\bibfnamefont {N.}~\bibnamefont {Luick}},
  \bibinfo {author} {\bibfnamefont {L.}~\bibnamefont {Sobirey}}, \bibinfo
  {author} {\bibfnamefont {J.}~\bibnamefont {Siegl}}, \bibinfo {author}
  {\bibfnamefont {T.}~\bibnamefont {Lompe}}, \bibinfo {author} {\bibfnamefont
  {H.}~\bibnamefont {Moritz}}, \bibinfo {author} {\bibfnamefont {L.~W.}\
  \bibnamefont {Clark}}, \ and\ \bibinfo {author} {\bibfnamefont
  {C.}~\bibnamefont {Chin}},\ }\href {\doibase 10.1364/OE.25.008670} {\bibfield
   {journal} {\bibinfo  {journal} {Opt. Express}\ }\textbf {\bibinfo {volume}
  {25}},\ \bibinfo {pages} {8670} (\bibinfo {year} {2017})}\BibitemShut
  {NoStop}%
\bibitem [{\citenamefont {Chekhova}\ \emph {et~al.}(2015)\citenamefont
  {Chekhova}, \citenamefont {Leuchs},\ and\ \citenamefont
  {Zukowski}}]{Chekhova2015}%
  \BibitemOpen
  \bibfield  {author} {\bibinfo {author} {\bibfnamefont {M.~V.}\ \bibnamefont
  {Chekhova}}, \bibinfo {author} {\bibfnamefont {G.}~\bibnamefont {Leuchs}}, \
  and\ \bibinfo {author} {\bibfnamefont {M.}~\bibnamefont {Zukowski}},\ }\href
  {\doibase 10.1016/j.optcom.2014.07.050} {\bibfield  {journal} {\bibinfo
  {journal} {Optics Communications}\ }\textbf {\bibinfo {volume} {337}},\
  \bibinfo {pages} {27} (\bibinfo {year} {2015})}\BibitemShut {NoStop}%
\bibitem [{\citenamefont {Fetter}\ and\ \citenamefont
  {Walecka}(2003)}]{Fetter2003}%
  \BibitemOpen
  \bibfield  {author} {\bibinfo {author} {\bibfnamefont {A.~L.}\ \bibnamefont
  {Fetter}}\ and\ \bibinfo {author} {\bibfnamefont {J.~D.}\ \bibnamefont
  {Walecka}},\ }\href@noop {} {\emph {\bibinfo {title} {{Quantum Theory of
  Many-Particle Systems}}}}\ (\bibinfo  {publisher} {Dover Publications},\
  \bibinfo {year} {2003})\ p.\ \bibinfo {pages} {640}\BibitemShut {NoStop}%
\end{thebibliography}

\newcommand{\beginsupplement}{%
        \setcounter{table}{0}
        \renewcommand{\thetable}{S\arabic{table}}%
        \setcounter{figure}{0}
        \renewcommand{\thefigure}{S\arabic{figure}}%
        \setcounter{equation}{0}
        \renewcommand{\theequation}{S\arabic{equation}}%
     }
\clearpage
     
\newpage
\onecolumngrid
\beginsupplement
\begin{center}
\large{\bf{Supplement Material for: 

{\textquotedblleft}Emission of Spin-correlated Matter-wave Jets from Spinor Bose-Einstein Condensates{\textquotedblright} }}
\end{center}

\section{I. Emission of the matter-wave jets in Two dimensions}
\begin{figure*}[!h]
    \centering
    \includegraphics[width=0.8\linewidth]{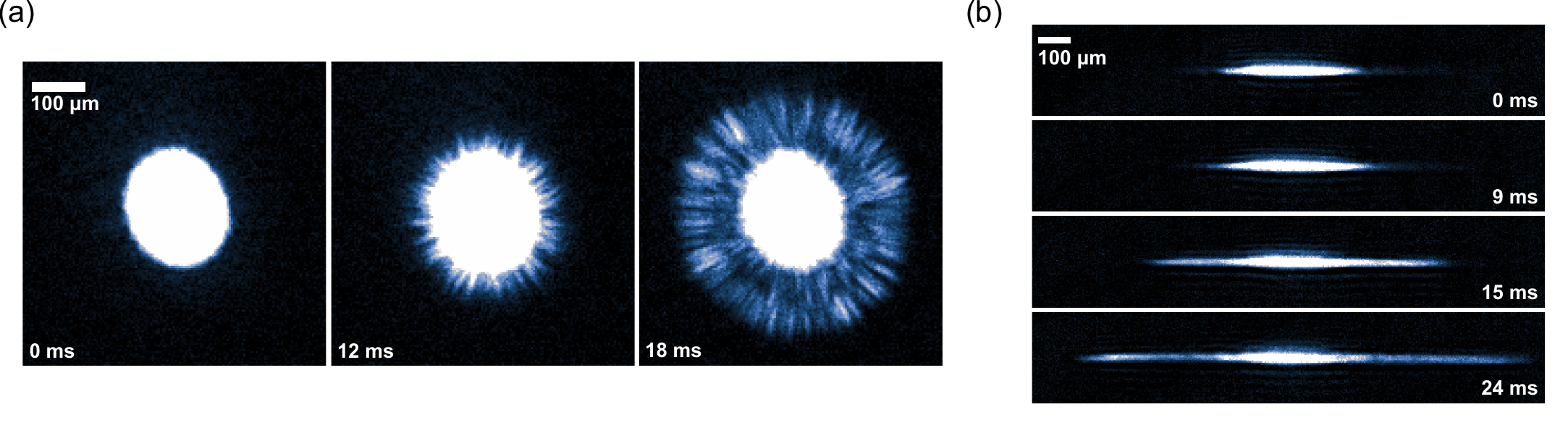}
    \caption{\textbf{\emph{in-situ} image of the matter-wave jets.}     
    (a) Top and (b) side images during the matter-wave jet emission under the quadratic Zeeman energy $q=-1.8$~kHz and $q=3.4$~kHz, respectively. All images are from a single realization of the experiment, and the hold time is noted at the bottom of each image. 
    }
    \label{fig:insitu}
\end{figure*}

Like the super-radiant light scattering experiment, the directionality of the matter-wave jet is determined by the geometry of the condensates. This is because the gain factor for the collision amplification process is given by the wavefunction overlap between the source condensate and scattered pair state~\cite{Pu2000,Vardi2002}. When the condensate size is much larger than the characteristic wavelength of the pair state, a sufficiently large gain can be obtained, displaying matter-wave amplification. In our experiment, the condensates are prepared in two dimensions since the chemical potential ($320$~Hz) is about half of the trapping frequency along the vertical direction ($\omega_z=2\pi\times680$~Hz). The axial motions of the condensates are restricted to the harmonic ground state with the spatial extent of $l_z=\sqrt{\hbar/M\omega}=1.5~\mu$m. In the $xy$-plane, it roughly follows the Thomas-Fermi profile with $R_{TF}\simeq 100~\mu$m. The generated atom pairs after spin-mixing collisions have a kinetic energy of few kHz, which have a characteristic wavelength of $\lambda_{mw}\sim4~\mu$m. In such parameters, we satisfy the following conditions, $l_z<\lambda_{mw} \ll R_{TF}$, and therefore, the collisional amplification processes are dominantly occurring in the planar direction (In FIG. \ref{fig:insitu}).

\section{II. Calibration of quadratic Zeeman energy}
To characterize the quadratic Zeeman energy $q$, we first calibrate the external bias field $B$ from the Rabi spectroscopy on the $\ket{F=1,m_F= 1}\rightarrow\ket{F=2,m_F=0}$ state transition. The field uncertainty is about 0.2~mG at 1~G of the magnetic field, and the quadratic Zeeman energy is determined by $q=h\times 610\ \text{Hz/G}^2\times B^2$. The energy scale will be compared to the kinetic energy of matter-wave jets of spin $\ket{0}$ state in the next section.

For the matter-wave jets of opposite spin states, $\ket{F=1,m_F=1}(\ket{\uparrow})$ and $\ket{F=1,m_F=-1}(\ket{\downarrow})$, we apply a microwave dressing field to the atoms through a single-loop coil inside of our vacuum chamber so that the quadratic Zeeman energy can be tuned to have negative value ($q<0$)~\cite{Gerbier2006}. The microwave field is red-detuned to the $\ket{1, -1}\leftrightarrow \ket{2, -2}$ transition with a detuning $\delta_d$ under the bias field $B=687.7(1)$ mG. This scheme minimizes the population in the $F=2$ state for a given energy shift because it provides sufficient detuning from the other transition lines such as $\ket{1, 0}\leftrightarrow\ket{2, -1}$. 

\begin{figure}[!h]
    \centering
    \includegraphics[width=0.8\columnwidth]{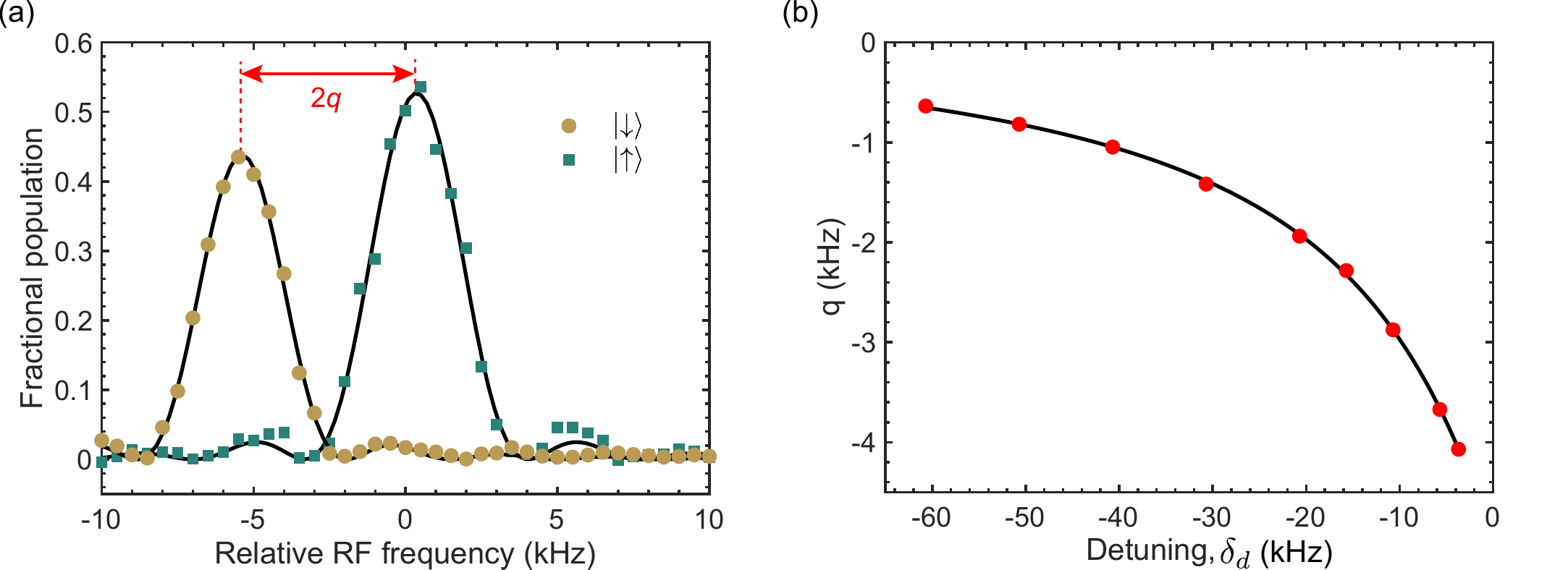}
    \caption{\textbf{Calibration of negative quadratic Zeeman energy.}
    (a) 
    Typical spectrum of the Zeeman sub-level spectroscopy. 
    We probe $\ket{1, 0}\rightarrow \ket{1, \pm1}$ transitions while the dressing field is on.
    The solid lines are fitted curve.
    Half of the frequency difference between the two transitions corresponds to the quadratic Zeeman energy, $q$.
    (b) 
    We interpolate the data (the distance between two spectrums in (a)) to calibrate $q$ as a function of the microwave frequency, $f$.
    The solid line is the fitted curve we used for the interpolation.
    See the text for the expression.
    The experiments are mostly performed around $q=-2$ kHz.
    }
    \label{fig:QZEcal}
\end{figure}

Probing the Zeeman sub-level transitions ($\ket{0}\rightarrow \ket{\uparrow}, \ket{\downarrow}$) under the microwave field, we directly measure the magnitude of $q$, instead of characterizing all experimental parameters, such as microwave polarization, detuning, and power. FIG.~\ref{fig:QZEcal}(a) shows the frequency difference between the two transitions, which corresponds to $2q$, and we control the $q$ by changing the detuning [FIG.~\ref{fig:QZEcal}(b)]. The measurements can be well described by an expression, $q= q_{0} + \left(f-f_0 - \sqrt{(f-f_0)^2+\Omega^2} \right)/2$, where $f$ is the microwave frequency, and $q_0, f_0$, and  $\Omega$ are the characteristic fit parameters of the dressing scheme. The measurement uncertainty of $q$ under the dressing field is less than $100$~Hz.

\newpage
\section{III. Kinetic energy and quadratic Zeeman energy}
\begin{figure}[!h]
    \centering
    \includegraphics[width=0.5\columnwidth]{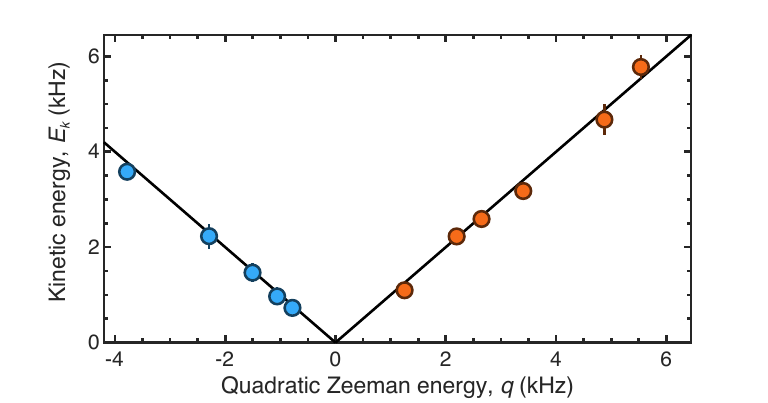}
    \caption{\textbf{Kinetic energy of the jets.} 
    By varying the free expansion time, we measured the kinetic energy of the emitted atom.
    The solid line is $E_k= |q|$, representing the kinetic energy source is the excess internal Zeeman energy.
    The error bars show the uncertainty (1-$\sigma$) of the momentum fit.
    The color code is for the sign of $q$, and we used a different quench protocol for each sign (see the main text).
    We keep the hold time short (about 10\% of the harmonic period) in order to reduce the effect of trapping potential.
    }
    \label{fig:QvsKE}
\end{figure}

The radial momentum of the atomic beam is measured by taking images with various expansion time (time-of-flight, TOF), and we calculate its average kinetic energy in radial direction ($E_k$). The $E_k$ is plotted as a function of the quadratic Zeeman energy in FIG. \ref{fig:QvsKE}, which is well described by $E_k\simeq |q|$ and is consistent with the Bogoliubov theory ($E_k\simeq |q|-|c|$)~\cite{Kawaguchi2012}. The small spin-dependent interaction energy, $c=-h\times160$~Hz (about 10$\%$ of $q$), is hindered by the measurement uncertainties such as limited TOF, trap curvature, and the finite size of the jets. This results indicate that the kinetic energy of the jets for all spin states are originated from the internal Zeeman energy.

\newpage
\section{IV. Thermal-like distribution of the jets}
\begin{figure}[!h]
    \centering
    \includegraphics[width=0.45\columnwidth]{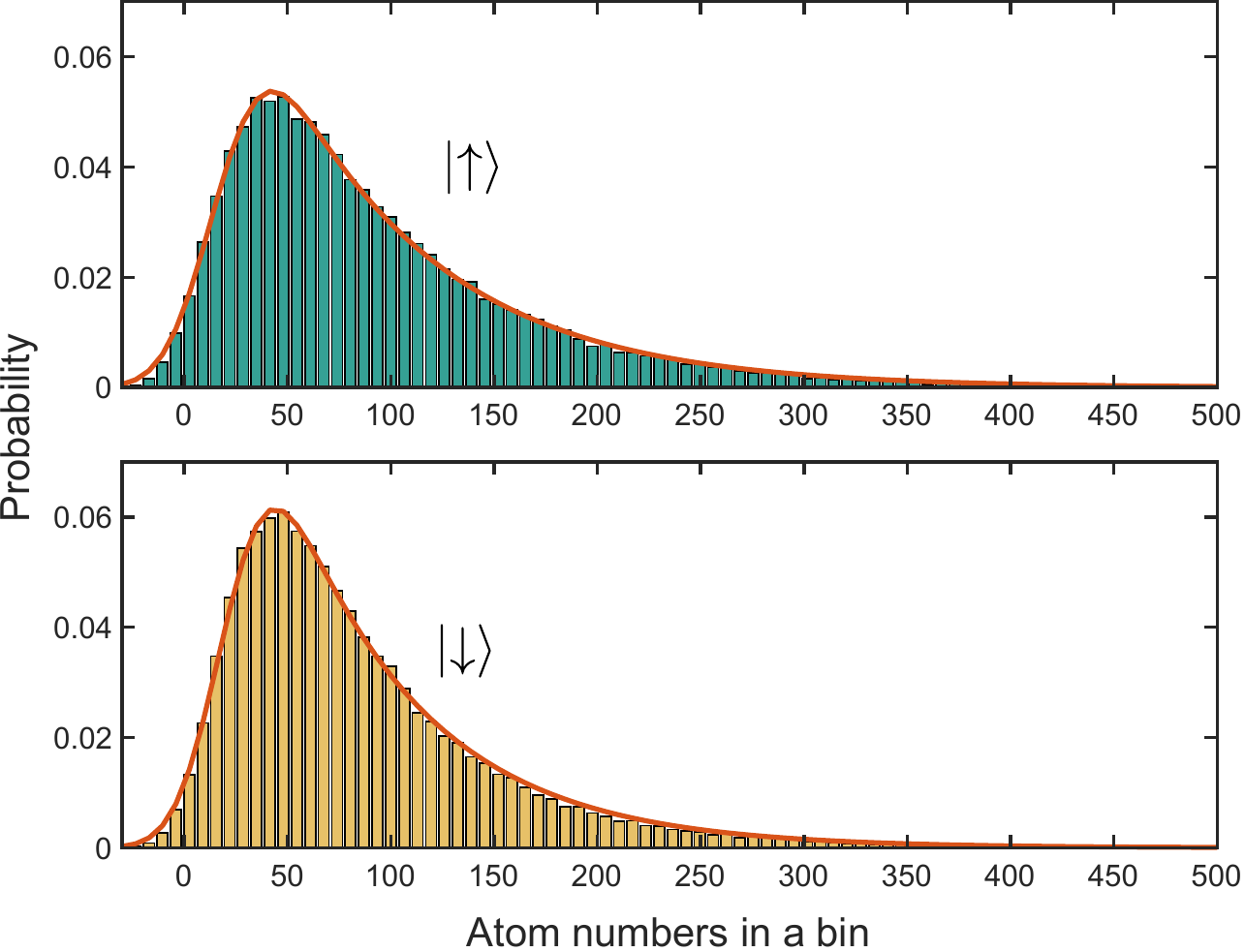}
    \caption{
    \textbf{Thermal like distribution of the jets.} 
    The atom number is sampled with the angular bin size of 1$^\circ$. 
    Solid lines are fitted curves, an exponentially decaying function convolved with a Gaussian.
    We used the same data as FIG. 3(b) in the main text. 
     }
    \label{fig:histogram}
\end{figure}

Figure \ref{fig:histogram} shows the histogram of the atom number for each spin state in the bin size of about 1$^\circ$. The distribution of each spin state is characterized by an exponentially decaying function convolved with a Gaussian function (accounting the detection noise), which is the thermal-like distribution as pointed out in \cite{Hu2019, Mias2008}. The slight difference in the distributions is attributed to the atom loss in $\ket{\downarrow}$ state because of the microwave dressing, and its details are explained in the following section.

\section{V. Discussion on the separability analysis}
In this section, we discuss our experimental imperfections that could increase entanglement witness. We adopt the separability criterion of the collective spin vectors that have been developed for bright squeezed vacuum state in quantum optics~\cite{Iskhakov2012}. 
\begin{eqnarray} \label{eq:witness}
	 \textsf{W} &=&\sum_{i=x, y, z} \frac{\text{Var}\qty(\hat{\text{J}}_i)}{2\ev{\hat{\text{J}}_0}}\geq 1,\\
	\nonumber
        \ev{\hat{\text{J}}_0} &\equiv& \ev{\sum_{m=\uparrow,\downarrow}[N_m(\phi) + N_m(\phi+\pi) ]},\\ 
        \nonumber
        \hat{\text{J}}_i &\equiv& \hat{P}_i(\phi) \pm  \hat{P}_i(\phi+\pi),\quad \\ 
        \nonumber
        \hat{P}_i(\phi) &\equiv& \int^{\Delta\phi/2}_{-\Delta\phi/2}{ [ \hat{n}_{\uparrow i}(\phi+\phi') - \hat{n}_{\downarrow i}(\phi+\phi') ]d\phi' }.
\end{eqnarray} 
The index $i=(x,y,z)$ refers to the spin axis, and the evaluation of the spin vector depends of the target Bell states. That is, three triplet Bell pair states take a minus (plus) sign to calculate the transverse (longitudinal) spin length and the witness $\textsf{W}_{\textsf{triplet}}$. For the singlet Bell pair state $\ket{\Psi_{S}}=\left( \hat{a}_{ \mathbf{k}, \uparrow}\hat{a}_{ \mathbf{-k}, \downarrow}  - \hat{a}_{ \mathbf{k}, \downarrow}\hat{a}_{ \mathbf{-k}, \uparrow} \right)^{\dagger}\ket{\text{vac}}$, the witness $\textsf{W}_{\textsf{singlet}}$ should be calculated with minus sign in all ($x$,$y$, and $z$) directions of the spin vectors. Investigating the separability criteria for both quantum states, we present three main sources that disturb entanglement certification besides the near field effect.

\begin{figure}[!h]
    \centering
    \includegraphics[width=0.9\columnwidth]{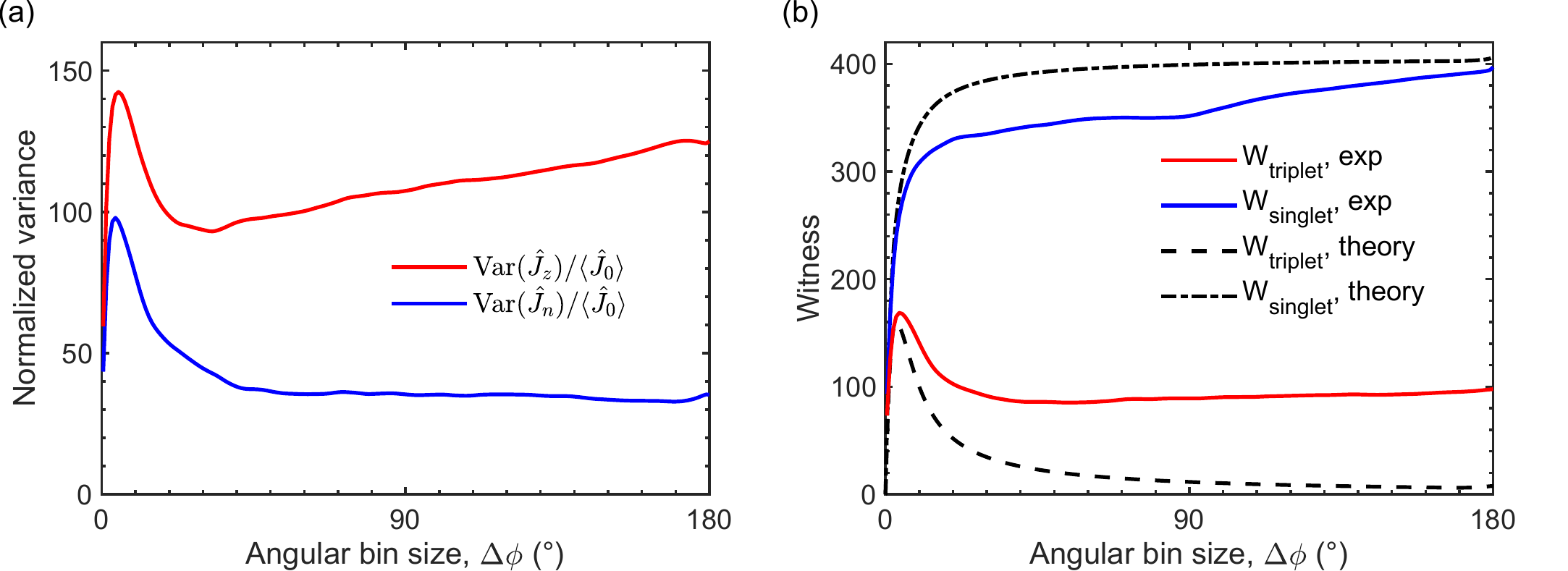}
    \caption{
        (a) Normalized variances of the collective spin vectors $\hat{\text{J}}_z$ and $\hat{\text{J}}_{n}$ as a function of bin size. The index $n$ stands for the normal to $z$-axis.
        We calculate the observables with two angular bins with a $180\ ^\circ$ difference (back-to-back propagating) as two modes for the separability test.
        The unbalanced loss leads quadratically increasing trend for $z$-directional collective spin variance (linear growth of the normalized variance). 
        (b) Witness with different target Bell configurations.
        We calculate the witnesses for different target states using the same data.
        As the state oscillates between singlet and triplet state, the calculated value of $\mathsf{W}_{\mathsf{triplet}}$ (or $\mathsf{W}_{\mathsf{singlet}}$) also oscillates in between the two theoretical lines.
             }
    \label{fig:witness} 
\end{figure}

\textbf{Unbalanced atom losses.} We observe the variance of longitudinal spin vector linearly increases as a function of the angular bin size (atom number), $\text{Var}(\hat{\text{J}}_z)\propto N_{\text{lost}}^{2}$ [FIG. \ref{fig:witness}(a)], which indicates spin selective atom losses~\cite{Simon2003}. Measuring the population imbalance (${I}=N_{\uparrow}-N_{\downarrow}$) after rotation of the spin axis, we observe the atom loss occurs in $\ket{\downarrow}$ state, which is estimated to about 20$\%$ during the hold time $t_h=7.5$~ms. We consider the atom are lost via inelastic collisions in the upper hyperfine state $\ket{2, -2}$ under the dressing field ~\cite{Weiner1999,Gerton1999}, where its fraction is measured  to 20$\%$ when we turn off the optical trap and reduce the atomic density.

\textbf{Detection uncertainty.} 
The calibration of atom numbers can be done by studying fluctuations of atom number difference in a coherent superposition state~\cite{Itano1993,Esteve2008,Reidel2010,Hamley2012}. This method allows one to count several hundreds of $^{87}$Rb atoms with few atom uncertainty even with absorption images~\cite{Muessel2013}. However, it is not directly applicable to $^7$Li atoms because of its light mass (and large doppler shift during imaging) and unresolved level structures of the $D_2$ transition lines~\cite{Hueck2017}. With our best efforts, the fluctuations are minimized at low probe beam intensity ($\simeq 0.2I_{sat}$, $I_{sat} = 2.5$ mW/cm$^2$ is the saturation intensity), where the variance of number imbalance for the superposition state  $[(\ket{\uparrow}-\ket{\downarrow})/\sqrt{2}]^{\otimes N}$ is 100 times larger than the total particle number: $\text{Var}(N_{\uparrow}-N_{\downarrow})\sim100\times({N_{\uparrow}+N_{\downarrow}})$. The contribution of the detection uncertainty in the entanglement witness would be around $\sim50$ at the most [FIG. \ref{fig:witness}(a)].

\textbf{Triplet-singlet mixing from the field gradient.} 
Under inhomogeneous magnetic field, the triplet Bell state $\ket{\Psi_{T}}=\left(\hat{a}_{\mathbf{k}, \uparrow}\hat{a}_{\mathbf{-k},\downarrow}+\hat{a}_{\mathbf{k},\downarrow}\hat{a}_{\mathbf{-k},\uparrow} \right)^{\dagger}\ket{\text{vac}}$ can be transformed into the singlet Bell state $\ket{\Psi_S}$ after a hold time because of different phase accumulation speed in the spin-momentum pair states. The field gradient in our setup is estimated to $4$ mG/cm, which leads to populate $\sim 30\%$ (maximal) of the singlet component in the matter-wave jets during the 7.5~ms of hold time. This could increase the entanglement witness for $\textsf{W}_{\textsf{triplet}}$ and reduce $\textsf{W}_{\textsf{singlet}}$ for the singlet Bell state [FIG. \ref{fig:witness}(b)].

\newpage
\section{VI. Theoretical model}
We employ the Bogoliubov theory of inhomogeneous spinor condensate \cite{Kawaguchi2012,Wu2019} to understand the experiment.  We try to follow the notations introduced in the Ref.~\cite{Kawaguchi2012}. Here, $\{+, 0, -\}$ denotes $m_F = \{+1, 0, -1\}$ states of $F=1$ manifold. We start from the second quantized Hamiltonian, $\hat{H} = \hat{H}_0 + \hat{V}$, where the non-interacting part is given as
\begin{equation*}
    \begin{split}
        \hat{H}_0 &= \int\mathrm{d}\mathbf{r} \sum_{m, m' \in \{+, 0, -\}} { \hat{\psi}^{\dagger}_{m}(\mathbf{r})\left[-\frac{\hbar^2\nabla^2}{2M} + U_{\text{trap}}(\mathbf{r})+ qm^2\delta_{mm'}\right]\hat{\psi}_{m'}(\mathbf{r}) } \\
        &\equiv  \int\mathrm{d}\mathbf{r} \sum_{m, m' \in \{+, 0, -\}} { \hat{\psi}^{\dagger}_{m}(\mathbf{r})\hat{h}_m (\mathbf{r})\hat{\psi}_{m'}(\mathbf{r}) }.
    \end{split}
\end{equation*}
$\hat{\psi}_{m}$ is the Bose field operator, $M$ is the mass of the atom, $U_{\text{trap}}$ is the harmonic trapping potential, and $q$ is the quadratic Zeeman energy. The interaction part is 
\begin{align*}
        \hat{V} & =  \frac{1}{2}\int\mathrm{d}\mathbf{r}  c_0 :\hat{n}^2(\mathbf{r}) : + c_2 : \hat{\mathbf{F}}^2(\mathbf{r}) : \\
        & = \frac{1}{2}\int\mathrm{d}\mathbf{r}     
        \begin{pmatrix}
            \hat{n}_{-} & \hat{n}_{0} & \hat{n}_{+}
        \end{pmatrix}
        \begin{pmatrix}
            g_2 & g_2  & c_0 - c_2  \\
            g_2 &  c_0 & g_2  \\
            c_0 - c_2 & g_2  & g_2  \\ 
        \end{pmatrix}
        \begin{pmatrix}
            \hat{n}_{-} \\ \hat{n}_{0} \\ \hat{n}_{+}
        \end{pmatrix}
        + 2c_2 ( \hat{\psi}^{\dagger}_{-}\hat{\psi}^{\dagger}_{+}\hat{\psi}^{}_{0}\hat{\psi}^{}_{0} + \hat{\psi}^{\dagger}_{0}\hat{\psi}^{\dagger}_{0}\hat{\psi}^{}_{+}\hat{\psi}^{}_{-} ).
\end{align*}
$c_0 = \frac{g_0+2g_2}{3}$ ($c_2 = \frac{g_2 - g_0}{3}$) is the spin independent (dependent) interaction coefficient. $g_2 = \frac{4\pi\hbar^2}{M}a_2$ ($g_0= \frac{4\pi\hbar^2}{M}a_0$), $a_2 = 7a_B$ ($a_0 = 25a_B$) for \textsuperscript{7}Li, where $a_B$ is the Bohr radius \cite{Huh2020}. The colons denote normal ordering. $\hat{n} = \sum_{m} \hat{n}_{m} = \sum_{m}{ \hat{\psi}_{m}^{\dagger}(\mathbf{r})\hat{\psi}_{m}(\mathbf{r}) }$ is the density operator, $\hat{F}_{\nu} = \sum_{m, m'} { \hat{\psi}_{m}^{\dagger}(\mathbf{r}) (\mathsf{f}_{\nu})_{mm'} \hat{\psi}_{m'}(\mathbf{r}) }$ with spin-1 matrices,
\begin{equation*}
    \mathsf{f}_{x} = \frac{1}{\sqrt{2}} 
    \begin{pmatrix}
        0 & 1 & 0 \\
        1 & 0 & 1 \\
        0 & 1 & 0 \\
    \end{pmatrix}, \quad
    \mathsf{f}_{y} = \frac{i}{\sqrt{2}} 
    \begin{pmatrix}
            0 & -1  & 0  \\
            1 &  0  & -1 \\
            0 &  1  & 0  \\
    \end{pmatrix}, \quad
    \mathsf{f}_{z} = 
    \begin{pmatrix}
            1 &  0  & 0  \\
            0 &  0  & 0  \\
            0 &  0  & -1 \\
    \end{pmatrix}.
\end{equation*}
The first term describes the density-density interaction. The second term corresponds to the spin-mixing channel relevant for the jet formation. 

\subsection{Analytic expression for homogeneous system}
\begin{figure}[!h] 
    \centering
    \includegraphics[width=0.4\columnwidth]{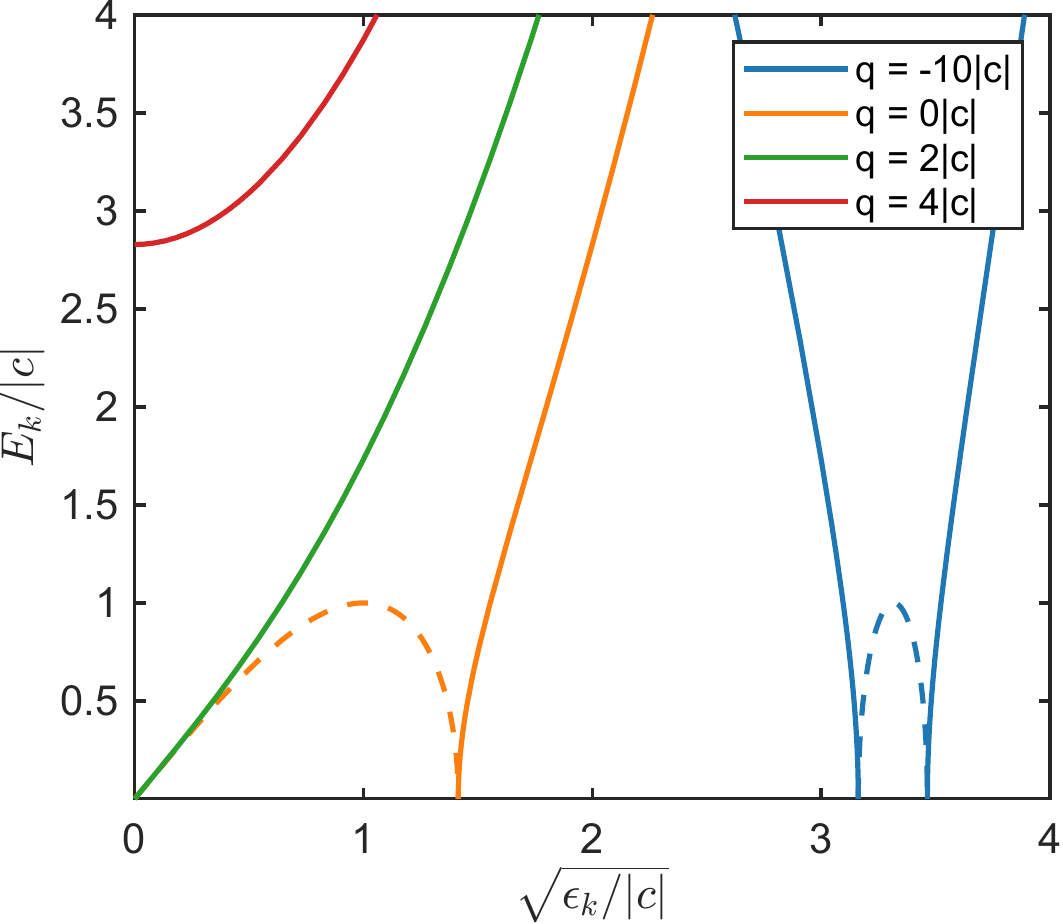}
    \caption{
    \textbf{Bogoliubov dispersion. }
    Solid (dashed) lines are the real (imaginary) part of $E_k$.  
    For $q$ larger than $2|c|$, we see the excitation gap. 
    As $q$ decreases, the gap decreases and the instability (imaginary energy) emerges. 
    The instability dome gets narrower as the magnitude of the $q$ increases. The width of the dome is simply $2|c|$ for $|q|\gg |c|$ regime. } 
    \label{dispersion}
\end{figure}

In a homogeneous system, we can obtain an analytic expression of matter-wave jets. The procedure is similar to \cite{Kawaguchi2012, Mias2008}. We expand the field operator with the plain wave basis.
\[
    \hat{\psi}_{m}(\mathbf{r}) = \frac{1}{\sqrt\Omega}\sum_{\mathbf{k}}{\hat{a}_{\mathbf{k}, m} e^{i\mathbf{k\cdot r}}},
\]
where $\Omega$ is for the normalization.
When the mean field part is $\boldsymbol{\Phi} = (0, 1, 0)^T$, the Bogoliubov Hamiltonian can be written as follows.
\begin{equation*}
    \hat{H}^{B} = \sum_{ \mathbf{k} \neq \mathbf{0} , m} \mathsf{Re}(E_{\mathbf{k}, m}) \left(\hat{b}^{\dagger}_{\mathbf{k}, m} \hat{b}_{\mathbf{k}, m} + \frac{1}{2}\right) +\sum_{ \mathbf{k} \neq \mathbf{0} , m} ' \mathsf{Im}(E_{\mathbf{k}, m}) \left(\hat{b}_{\mathbf{k}, m} \hat{b}_{\mathbf{-k}, m} + \hat{b}^{\dagger}_{\mathbf{k}, m} \hat{b}^{\dagger}_{\mathbf{-k}, m}\right),
\end{equation*}
where 
\begin{align*}
    &E_{\mathbf{k}, \pm } = \sqrt{(\epsilon_{\mathbf{k}}+q) (\epsilon_{\mathbf{k}}+q +2c_2 n)} \\
    &\hat{b}_{\mathbf{k}, \pm } = -\sqrt{\frac{\epsilon_{\mathbf{k}}+q +c_2 n + E_{\mathbf{k}, \pm }}{2E_{\mathbf{k}, \pm }}}\hat{a}_{\mathbf{k}, \pm } + \sqrt{\frac{\epsilon_{\mathbf{k}}+q +c_2 n - E_{\mathbf{k}, \pm }}{2E_{\mathbf{k}, \pm }}}\hat{a}^{\dagger}_{\mathbf{-k}, \mp }. \label{eq:firstBT}
\end{align*}
$E_{\mathbf{k}}$ is the complex energy eigenvalue, $\epsilon_{\mathbf{k}} \equiv \hbar^2k^2/2M$ is the single particle kinetic energy. The prime symbol $(')$ reminds not to double-count $\mathbf{k}$. The second term represents the instability that is in our main interest. The energy eigenvalues for different $q$ with a shorthand, $c \equiv c_2n$, is presented in FIG. \ref{dispersion}.
 
For high quadratic Zeeman energy, $|q| \gg |c|$, instability sharply peaks at $\epsilon_{\mathbf{k}} = |q|-|c| \approx |q|$. The energy eigenvalue becomes pure imaginary at this point and Hamiltonian is further reduced to the following form.
\begin{equation*}
    \hat{H}^{B} = \sum_{ \epsilon_{\mathbf{k}} \approx |q|} ' c \left( \hat{a}^{\dagger}_{ \mathbf{-k}, -}\hat{a}^{\dagger}_{ \mathbf{k}, +}  + \hat{a}^{\dagger}_{ \mathbf{-k}, +}\hat{a}^{\dagger}_{ \mathbf{k}, -} \right) + \text{H.c}.
\end{equation*}
Neglecting the condensate depletion effect, this Hamiltonian is time-independent.  We obtain the equation of motion with the aid of SU(1,1) algebra. The final state of the jet, $\ket{f(t)}$ is
\begin{equation*} 
    \ket{f(t)} = \exp{-i\hat{H}^{B} t/\hbar }\ket{\text{vac}} = \bigotimes _{ \epsilon_{\mathbf{k}} \approx |q| }^{'} \sum_{j=0}^{\infty}\lambda_{j}(t)\left[(\hat{a}_{ \mathbf{-k}, -}\hat{a}_{ \mathbf{k}, +} + \hat{a}_{ \mathbf{-k}, +}\hat{a}_{ \mathbf{k}, -})^{\dagger}\right]^j \ket{\text{vac}},
\end{equation*}
where $\lambda_{j}(t) = \frac{{(-i\tanh{(|c|t)})^j}}{j!\cosh^2{(|c|t)}}$. Note that this state is analogous to the (one of the triplet form) bright squeezed vacuum states \cite{Chekhova2015}.

\subsection{Time-dependent Bogoliubov theory}
We perform numerical calculations that include experimental situations, such as the system's finite size, condensate depletion, and harmonic curvature. Here, we summarize the setup for the numerical calculations we showed in the main text. In order to describe the dynamics of the system, we consider a grand canonical ensemble with the potential, $\hat{K} = \hat{H} - \mu \hat{N}$, where $\mu$ is the chemical potential and $\hat{N}$ is the total particle number operator. We also apply the Bogoliubov prescription: $ \hat{\psi}_{m} (\mathbf{r}, t) \rightarrow e^{i\hat{K}t} ( {\Phi}_{m} (\mathbf{r}) +  \hat{\varphi}_{m} (\mathbf{r}, 0) ) e^{-i\hat{K}t} $, ${\Phi}_{m}$ is the mean field stationary solution that describe the condensates and $\hat{\varphi}_{m}$ is the non-condensates operator. We obtain ${\Phi}_{m} (\mathbf{r})$ and $\mu$ using Gross-Pitaevskii equation under Thomas-Fermi approximation. The equation of motion, up to the first order of $\hat{\varphi}$, is \cite{Fetter2003,Wu2019}
\begin{equation*}
    i\hbar \frac{\partial}{\partial t} \hat{\varphi}_m(\mathbf{r}, t) = \qty( \hat{h}_m(\mathbf{r}) - \mu ) \hat{\varphi}_m
    + \sum_{m_2, m'_1, m'_2} \qty(C^{m m'_1}_{m_2 m'_2} + C^{m m'_2}_{m'_1 m_2})  \Phi^*_{m'_2}\Phi_{m'_1}\hat{\varphi}_{m_2} 
    + C^{m m_2}_{m'_1 m'_2}\Phi_{m'_2}\Phi_{m'_1} \hat{\varphi}^{\dagger}_{m_2},
\end{equation*}
where
\begin{equation*}
    C^{m_1 m_2}_{m'_1 m'_2} = c_0 \delta_{m_1 m'_1} \delta_{m_2 m'_2} + c_2 \sum_{\nu = x, y, z}{(\mathsf{f}_\nu)_{m_1 m'_1} (\mathsf{f}_\nu)_{m_2 m'_2}}.
\end{equation*}
We neglect the excitations of spin-0 component (spin-$\pm$1 component) for $q < 0$ ($q > 0$) case. 

\begin{figure}[!h]
    \centering
    \includegraphics[width=0.4\columnwidth]{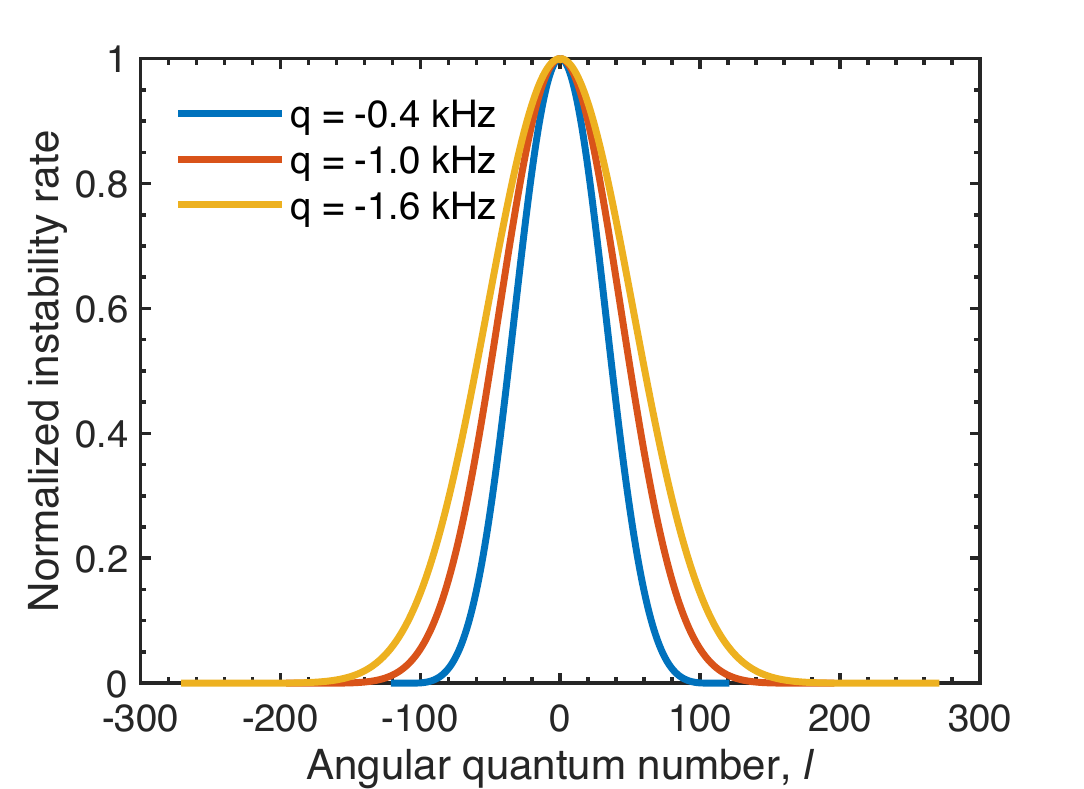}
    \caption{
        \textbf{Instability for different angular mode.}
        Normalized instability rate as a function of the angular quantum number, $l$. 
        As the internal energy ($q$) increases, the distribution expand because of the increased density of states for higher energy.}
    \label{fig:mode}
\end{figure}

We continue with the $q<0$ case ($q>0$ case is analogous to this procedure).  We apply the Bogoliubov transformation and expand the coefficients in terms of the eigenfunctions.
\begin{equation*}
    \begin{split}
        \hat{\varphi}_+ = \sum_{l, n_r} { \qty( u_{+,l, n_r}(\mathbf{r}, t) \hat{\alpha}_{n_r , l}+ v_{+,l, n_r}^{*}(\mathbf{r}, t) \hat{\beta}_{l, n_r}^{\dagger} ) }, \\
        \hat{\varphi}_- = \sum_{l, n_r} { \qty( u_{-,l, n_r}(\mathbf{r}, t) \hat{\beta}_{l, n_r} + v_{-,l, n_r}^{*}(\mathbf{r}, t) \hat{\alpha}_{l, n_r}^{\dagger} ) }.
    \end{split}
\end{equation*}
$u$ and $v$ are the eigenfunction for single-particle Hamiltonian, $\hat{h}_m$. $\hat{\alpha}$ and $\hat{\beta}$ satisfy the bosonic commutation relation.  The indices $(l, n_r)$ is for the quantum numbers of the eigenfunctions. Because the system is in 2D regime (chemical potential is half of the tight trapping frequency), dynamics can be neglected along the tight confinement direction. We use 2D polar coordinates, $\mathbf{r} = (\rho, \theta)$ and assume the trapping potential to be a simple harmonic oscillator. We assumed that the system is perfectly symmetric (with $8$ Hz harmonic frequency); we neglect the effect from the slight ellipticity of the system.  The resulting quantum numbers are $(l, n_r)$ for angular and radial quantum number with the following eigenfunction, $\phi_{n_r}^{l}$.
\begin{gather*}
    \phi_{n_r}^{l} (\mathbf{r}) = R_{n_r}^{l}(r) \frac{e^{il\theta}}{\sqrt{2\pi}},  \\
    R_{n_r}^{l}(r) = \sqrt{\frac{n_r !}{(n_r + l)!} \frac{k^2 \alpha^{l+1}}{2^l}}e^{-\alpha x^2/4}x^lL^l_{n_r}\left(\frac{\alpha x^2}{2}\right), \quad \alpha = \frac{1}{2n_r + l + 1}. 
\end{gather*}
Here, $L^l_{n_r}$ is the associated Laguerre polynomial and $x = k \cdot r= \sqrt{2M\omega(2n_r+l+1)/\hbar} \cdot r$. 
We can write $u$ and $v$ with the coefficients $U$ and $V$.
\begin{gather*}
    u_{m, l, n_r} = \sum_{n'_r}U^{l}_{m, n_rn'_r}\phi_{n'_r}^{l} (\mathbf{r}), \quad v_{m, l, n_r} = \sum_{n'_r}V^{l}_{m, n_rn'_r}\phi_{n'_r}^{l} (\mathbf{r})
\end{gather*}
Due to the symmetry of the system, different $l$ modes are decoupled. We construct the matrix equation of $U$ and $V$ for each $l$ and evaluate the matrix form Hamiltonian using the orthonormality of the eigenfunction. Corresponding instability rate distributions for several $q$ can be found in FIG. \ref{fig:mode}.

In FIG. 4(c) on the main text, we include the effect of field gradient.  We add spin dependent phase term to $\hat{h}_m \rightarrow \hat{h}_m + pm$, where $p = h\times$3 Hz. This prescription accounts averaged gradient effect. The exact consideration of the field gradient require calculations in 2D Cartesian coordinate.

We integrate the matrix equation to observe dynamics. We similarly consider the depletion to \cite{Wu2019}, i.e. subtracting the excited particle number from the condensate number for each time step. TOF is implemented using fast Fourier transform by converting the coordinate to Cartesian. In order to describe the reduced growth rate as presented in FIG. 2-(a) of the main text, we introduce a fitting parameter, $\beta$ for the interaction strength $c_2 \rightarrow \beta c_2$ and $\beta = 0.5$.

\end{document}